\documentclass[ALICE,manyauthors]{cernphprep}

\usepackage{graphicx}
\usepackage{epstopdf}
\usepackage{amssymb}
\usepackage{lineno}
\usepackage{xspace}
\usepackage{amsmath}
\usepackage{upgreek}
\usepackage{color}
\usepackage{hyperref}
\usepackage{doi}

\DeclareMathSymbol{\Gamma}{\mathalpha}{letters}{"00}

\newcommand{\krl}{\ensuremath{\kern-0.18em}}
\newcommand{\krr}{\ensuremath{\kern-0.09em}}
\newcommand{\tms}{\ensuremath{\kern-0.1em\times\kern-0.2em}}

\newcommand{\ptt}{\ensuremath{p_{\mathrm{T}}}\xspace}
\newcommand{\mT}{\ensuremath{m_{\mathrm{T}}}\xspace}
\newcommand{\pb}{\mbox{Pb--Pb}\xspace}
\newcommand{\au}{\mbox{Au--Au}\xspace}
\newcommand{\cu}{\mbox{Cu--Cu}\xspace}
\newcommand{\dau}{\mbox{d--Au}\xspace}
\newcommand{\ada}{\mbox{A--A}\xspace}

\newcommand{\rs}[1][7~TeV]{\ensuremath{\sqrt{s}=}~#1\xspace}

\newcommand{\rsnn}[1][2.76~TeV]{\mbox{\ensuremath{\sqrt{s_{\mathrm{NN}}}=}~#1}\xspace}
\newcommand{\rsnno}{\ensuremath{\sqrt{s_{\mathrm{NN}}}}\xspace}

\newcommand{\gvc}{\ensuremath{\mathrm{GeV}\krl/\krr c}\xspace}
\newcommand{\gvcc}{\ensuremath{\mathrm{GeV}\krl/\krr c^{2}}\xspace}
\newcommand{\mvc}{\ensuremath{\mathrm{MeV}\krl/\krr c}\xspace}
\newcommand{\mvcc}{\ensuremath{\mathrm{MeV}\krl/\krr c^{2}}\xspace}

\newcommand{\pion}{\ensuremath{\uppi}\xspace}
\newcommand{\pix}{\ensuremath{\pion^{\pm}}\xspace}
\newcommand{\pim}{\ensuremath{\pion^{-}}\xspace}
\newcommand{\pip}{\ensuremath{\pion^{+}}\xspace}

\newcommand{\kx}{\ensuremath{\mathrm{K}^{\pm}}\xspace}
\newcommand{\km}{\ensuremath{\mathrm{K}^{-}}\xspace}
\newcommand{\kp}{\ensuremath{\mathrm{K}^{+}}\xspace}

\newcommand{\ppi}{\ensuremath{\mathrm{p}\kern-0.05em/\krr\pion}\xspace}
\newcommand{\pxpix}{\ensuremath{(\mathrm{p}+\bar{\mathrm{p}})\krr/\krr(\pim+\pip)}\xspace}

\newcommand{\ks}{\ensuremath{\mathrm{K^{*0}}}\xspace}

\newcommand{\ksm}{\ensuremath{\mathrm{K^{*}\krr(892)^{0}}}\xspace}
\newcommand{\ksbm}{\ensuremath{\mathrm{\overline{K}^{*}(892)^{0}}}\xspace}
\newcommand{\ph}{\ensuremath{\upphi}\xspace}
\newcommand{\phm}{\ensuremath{\ph(1020)}\xspace}

\newcommand{\ksk}{\ensuremath{\ks\krl/\mathrm{K}}\xspace}

\newcommand{\kskm}{\ensuremath{\ks\krl/\km}\xspace}

\newcommand{\pks}{\ensuremath{\mathrm{p}\kern-0.1em/\ks}\xspace}
\newcommand{\pksm}{\ensuremath{\mathrm{p}\kern-0.1em/\ksm}\xspace}

\newcommand{\phipi}{\ensuremath{\ph\krl/\krr\pion}\xspace}
\newcommand{\phimpi}{\ensuremath{\phm\krl/\krr\pion}\xspace}
\newcommand{\phipix}{\ensuremath{\ph\krl/\krr(\pim+\pip)}\xspace}

\newcommand{\phik}{\ensuremath{\ph\krl/\mathrm{K}}\xspace}

\newcommand{\phikm}{\ensuremath{\ph\krl/\km}\xspace}

\newcommand{\pphi}{\ensuremath{\mathrm{p}\kern-0.1em/\krl\ph}\xspace}
\newcommand{\pphim}{\ensuremath{\mathrm{p}\kern-0.1em/\krl\phm}\xspace}
\newcommand{\pxphi}{\ensuremath{(\mathrm{p}+\bar{\mathrm{p}})\krr/\krr\ph}\xspace}

\newcommand{\omphi}{\ensuremath{\Omega\kern-0.05em/\krr\ph}\xspace}
\newcommand{\omxphi}{\ensuremath{(\Omega^{-}+\overline{\Omega}^{+})\kern-0.05em/\krr\ph}\xspace}
\newcommand{\ommphi}{\ensuremath{\Omega^{-}\kern-0.05em/\krr \ph}\xspace}
\newcommand{\ompphi}{\ensuremath{\overline{\Omega}^{+}\kern-0.05em/\krr \ph}\xspace}
\newcommand{\omphim}{\ensuremath{\Omega\kern-0.05em/\krr\phm}\xspace}

\newcommand{\lpi}{\ensuremath{\Lambda/\krr\pion}\xspace}
\newcommand{\xipi}{\ensuremath{\Xi\kern-0.1em/\krr\pion}\xspace}
\newcommand{\ompi}{\ensuremath{\Omega\kern-0.05em/\krr\pion}\xspace}

\newcommand{\dd}{\ensuremath{\mathrm{d}}}
\newcommand{\mpt}{\ensuremath{\langle\ptt\rangle}\xspace}
\newcommand{\rcp}{\ensuremath{R_{\mathrm{CP}}}\xspace}
\newcommand{\dndy}{\ensuremath{\dd N\krl/\krr\dd y}\xspace}
\newcommand{\ddn}{\ensuremath{\dd^{2}N\krl/\krr(\dd\ptt\dd y)}\xspace}
\newcommand{\npart}{\ensuremath{\langle N_{\mathrm{part}}\rangle}\xspace}
\newcommand{\dnc}{\ensuremath{\dd N_{\mathrm{ch}}\kern-0.06em /\kern-0.13em\dd\eta}\xspace}
\newcommand{\dncr}{\ensuremath{(\dnc)^{1/3}}\xspace}

\newcommand{\dedx}{\ensuremath{\dd E\krl/\krr\dd x}\xspace}
\newcommand{\stpc}{\ensuremath{\sigma_{\mathrm{TPC}}}\xspace}

\newcommand{\cn}{\ensuremath{\chi^{2}\krl/\krr n_{\mathrm{dof}}}\xspace}

\newcommand{\effr}{\ensuremath{\varepsilon_{\mathrm{rec}}}\xspace}
\newcommand{\effp}{\ensuremath{\varepsilon_{\mathrm{PID}}}\xspace}

\newcommand{\mpik}{\ensuremath{m_{\pion\mathrm{K}}}\xspace}
\newcommand{\mkk}{\ensuremath{m_{\mathrm{KK}}}\xspace}

\begin{document}

\begin{titlepage}

\PHyear{2014}
\PHnumber{060} 
\PHdate{02 April}

\title{K$\boldsymbol{^{*}}$(892)$\boldsymbol{^{0}}$ and $\boldsymbol{\upphi}$(1020) production in \pb collisions at $\boldsymbol{\sqrt{s_{\mathrm{NN}}}}$~=~2.76~TeV}
\ShortTitle{\ksm and \phm in \pb collisions}

\Collaboration{ALICE Collaboration
         \thanks{See Appendix~\ref{app:collab} for the list of collaboration
                      members}}
\ShortAuthor{ALICE Collaboration}

\begin{abstract}
The yields of the \ksm and \phm resonances are measured in \pb collisions at \linebreak\rsnn through their hadronic decays using the ALICE detector.  The measurements are performed in multiple centrality intervals at mid-rapidity $(|y|<0.5)$ in the transverse-momentum ranges $0.3<\ptt<5$~\gvc for the \ksm and $0.5<\ptt<5$~\gvc for the \phm.  The yields of \ksm are suppressed in central \pb collisions with respect to pp and peripheral \pb collisions (perhaps due to re-scattering of its decay products in the hadronic medium), while the longer lived \phm meson is not suppressed.  These particles are also used as probes to study the mechanisms of particle production.  The shape of the \ptt distribution of the \phm meson, but not its yield, is reproduced fairly well by hydrodynamic models for central \pb collisions.  In central \pb collisions at low and intermediate \ptt, the \pphim ratio is flat in \ptt, while the \ppi and \phimpi ratios show a pronounced increase and have similar shapes to each other.  These results indicate that the shapes of the \ptt distributions of these particles in central \pb collisions are determined predominantly by the particle masses and radial flow.  Finally, \phm production in \pb collisions is enhanced, with respect to the yield in pp collisions and the yield of charged pions, by an amount similar to the $\Lambda$ and $\Xi$.


\end{abstract}
\end{titlepage}
\setcounter{page}{2}

\section{Introduction\label{sec:intro}}

Ultrarelativistic heavy-ion collisions are expected to produce a hot and dense state of matter, the quark-gluon plasma~\cite{Petreczky_ConfinementX,Borsanyi_2010b,Borsanyi_TC}.  At a critical temperature of $T_{\mathrm{c}}\approx 160$~MeV~\cite{Borsanyi_TC,Aoki_TC1,Aoki_TC2} a cross-over transition between the partonic (\textit{i.e.}, a system with deconfined quarks) and hadronic phases is expected to take place.  Statistical models~\cite{Cleymans_2006,Rafelski_2005,Petran_Rafelski_2013a,Petran_Rafelski_2013b,Andronic2009,Andronic2009_Erratum,AndronicQM2011,Becattini_2004,Becattini_2010} have been successfully applied to particle yields in order to estimate the values of the chemical freeze-out temperature and the baryochemical potential.  However, resonance yields may deviate from the values expected from thermal models due to hadronic processes (re-scattering and regeneration) that might change the reconstructible resonance yields even after chemical freeze-out.  Resonance yields may be regenerated through pseudo-elastic scattering, in which particles scatter through a resonance state [\textit{e.g.}, $\pim\kp\rightarrow\ksm\rightarrow\pim\kp$ and $\km\kp\rightarrow\phm\rightarrow\km\kp$]~\cite{Bleicher_Stoecker,Markert_thermal,Vogel_Bleicher}.  Pseudo-elastic scattering does not change the abundances of the scattered particles, but may increase the measured yield of the resonance state through which they scattered.  If a resonance has a short enough lifetime, it may decay during the hadronic phase and its decay products may undergo elastic or pseudo-elastic scatterings.  Information about the resonance may be lost if at least one of its decay products elastically scatters in the hadronic medium or undergoes pseudo-elastic scattering via a different resonance state [\textit{e.g.}, a pion from a \ksm decay scatters with another pion, $\pim\pip\rightarrow\uprho(770)^{0}\rightarrow\pim\pip$]~\cite{Bliecher_Aichelin}.  The net effect of pseudo-elastic scattering on the yield of a resonance will depend on whether regeneration of that resonance is outweighed by re-scattering of its decay products through other resonances.  In the case of the \ksm, the $\pion$K interaction cross section~\cite{Matison_Kpi} is smaller than the $\pion\pion$ cross section~\cite{Protopopescu_pipi}, so re-scattering may dominate and the measured \ksm yield may be smaller than the yield at chemical freeze-out.  Calculations using Ultrarelativistic Quantum Molecular Dynamics (UrQMD)~\cite{UrQMD_Bass,UrQMD} predict that both regeneration and re-scattering affect the resonance yields predominantly for transverse momenta $\ptt\lesssim2$~\gvc~\cite{Bleicher_Stoecker,Bliecher_Aichelin}.  The final reconstructible resonance yields depend on the chemical freeze-out temperature, the scattering cross sections of its decay products, and the timescale during which re-scattering and regeneration are active in the hadronic phase, \textit{i.e.}, the time between chemical and kinetic freeze-out.  The model described in~\cite{Markert_thermal,Torrieri_thermal,Torrieri_thermal_2001b,Torrieri_thermal_2001b_erratum} combines thermal-model calculations with re-scattering effects in the hadronic phase.  It predicts the ratios of \linebreak(\ptt-integrated) resonance yields to the yields of stable particles as a function of both the chemical \linebreak freeze-out temperature and the lifetime of the hadronic phase.  While this model was derived for a \linebreak Relativistic Heavy Ion Collider (RHIC) collision energy (\rsnn[130~GeV]), its predictions span a wide range of freeze-out temperatures and hadronic lifetimes and remain valid at Large Hadron Collider (LHC) energies.

Chiral symmetry is expected to be restored~\cite{Petreczky} above the chiral transition temperature; resonances that decay when chiral symmetry was at least partially restored are expected to exhibit mass shifts and/or width broadening~\cite{Brown_Rho,Rapp2009,Brodsky_chiral,Eletsky}.  Regeneration of resonances in the late hadronic phase increases the fraction of resonances with vacuum masses and widths and may inhibit the observation of the signatures of chiral symmetry restoration.  Since model calculations indicate that re-scattering and regeneration modify the resonance signal more strongly for $\ptt\lesssim 2$~\gvc, signatures of chiral symmetry may be difficult to observe in the case of low-\ptt resonances which are reconstructed via hadronic decays.

This article presents measurements of the \ksm, \ksbm, and \phm mesons performed in multiple centrality intervals for \pb collisions at \rsnn using the ALICE detector.  The focus here is on low and intermediate \ptt [$0.3<\ptt<5$~\gvc for the \ksm and $0.5<\ptt<5$~\gvc for the \phm] and the integrated yields; results for high \ptt will be presented in a future article.  All measurements of the \ksm and \ksbm are averaged and these mesons are collectively referred to as \ks.  The \phm meson is referred to as \ph.  The ALICE detector is described in Sec.~\ref{sec:alice}, with the emphasis on the sub-detectors used in this analysis.  The data-analysis procedure is described in Secs.~\ref{sec:pid}-\ref{sec:corr}.  Results, including resonance yields, masses, widths, mean transverse momenta, ratios to non-resonances, comparisons to predicted \ptt distributions, and the \ph enhancement ratio are presented in Sec.~\ref{sec:results}.

\section{ALICE Experiment\label{sec:alice}}

A comprehensive description of the ALICE detector can be found in~\cite{ALICE_detector}.  The main detector components used in this analysis are the V0 detector, the Inner Tracking System (ITS), and the Time Projection Chamber (TPC), which are located inside a 0.5~T solenoidal magnetic field.  The V0 detector~\cite{ALICE_VZERO} consists of two scintillator hodoscopes placed on either side of the interaction point covering the pseudorapidity ranges $-3.7<\eta<-1.7$ and $2.8<\eta<5.1$.  A combination of hits in the V0 detector and the two innermost layers of the ITS is used is used as a minimum-bias trigger for \pb collisions~\cite{ALICE_multiplicity}.  Collision centrality is determined by using the multiplicity measured in the V0 detector along with Glauber-model simulations to describe the multiplicity distribution as a function of the impact parameter~\cite{ALICE_multiplicity,ALICE_centrality}.  These simulations give \npart, the mean number of nucleons which participated in collisions in a given centrality interval.  The ITS is made up of six cylindrical layers of silicon detectors with radii between 3.9 and 43 cm from the beam axis, covering the full azimuth.  The pseudorapidity range $|\eta|<0.9$ is covered by all six layers, with some of the individual layers covering larger ranges in pseudorapidity.  The TPC~\cite{ALICE_TPC}, which is the main tracking detector, is a large cylindrical drift detector that covers the pseudorapidity range $|\eta|<0.9$ with full azimuthal acceptance.  Multi-wire proportional chambers with cathode pad readout are arranged in 159 pad rows located at the ends of the TPC.  Hits in the ITS and TPC are used to reconstruct charged particle tracks, which are used in the final determination of the primary collision vertex.  The position resolution for the primary vertex in both the longitudinal direction and the transverse plane is $\sim10\;\upmu\mathrm{m}$ for heavy-ion collisions.  The TPC is also used to identify particles through their \dedx (specific energy loss) in the TPC gas.  The value of \dedx is calculated using a truncated-mean procedure in which the average is evaluated using only the 60\% of points with the lowest \dedx values measured along a given track.  The measured \dedx is then compared to the expected \dedx for a given particle species using a Bethe-Bloch parametrization.  The deviation from the expected \dedx value is expressed in units of the energy-loss resolution \stpc, which is 5\% for isolated tracks and 6.5\% for central collisions~\cite{ALICE_piKp_PbPb}.  The TPC allows kaons to be distinguished from pions for momenta $p<0.7$~\gvc and (anti)protons to be distinguished from pions and kaons for $p<1$~\gvc (with a separation power of 2$\sigma$ in both cases).

\section{Event and Track Selection\label{sec:pid}}

The yields of \ks and \ph mesons are measured in about 13 million \pb collisions recorded in 2010 in the 0-90\% centrality interval.  The position of the primary vertex along the beam axis is required to be within 10~cm of the center of the ALICE detector.  The \ks and \ph mesons are identified by reconstruction of their respective hadronic decays: $\ks\kern-0.15em\rightarrow\kern-0.15em\pix\mathrm{K}^{\mp}$ (branching ratio 0.666) and $\ph\kern-0.15em\rightarrow\kern-0.15em\km\kp$ (branching ratio 0.489)~\cite{PDG}.  The lifetimes in the vacuum of the \ks and \ph are $4.16\pm0.05$~fm/$c$ and $46.3\pm0.4$~fm/$c$, respectively~\cite{PDG}.  High-quality tracks are selected by requiring at least 70 reconstructed TPC clusters out of a possible 159 and requiring that the $\chi^{2}$ per cluster of the reconstructed tracks be less than 4.  Track momenta and pseudorapidity are restricted to the ranges $\ptt>150$~\mvc and $|\eta|<0.8$, respectively.  To reduce the number of secondary particles from weak decays, each track is required to have at least one hit in the innermost layer of the ITS and a small distance of closest approach (DCA) to the primary vertex in the $xy$ plane: $\mathrm{DCA}_{xy}<(0.0182+0.035\ptt^{-1.01})$~cm.  The distance of closest approach in the $z$ direction is also restricted: $\mathrm{DCA}_{z}<2$~cm.  The $\mathrm{DCA}_{z}$ cut is wider not because of the vertex resolution (which is similar for the longitudinal and transverse directions), but because of the tracking resolution, which is less precise for the $z$ direction than the transverse plane.  This is because the positions of points in the Silicon Pixel Detector (the innermost part of the ITS) are determined more precisely in the $xy$ plane.  The wide $\mathrm{DCA}_{z}$ cut is intended to remove particles that are highly displaced from the vertex.  Finally, in the \ks and \ph analyses, pion and kaon tracks are required to be within $2\stpc$ of the expected \dedx values for each particle species.

\section{Signal Extraction\label{sec:signal}}

\begin{figure*}
\includegraphics[width=38pc]{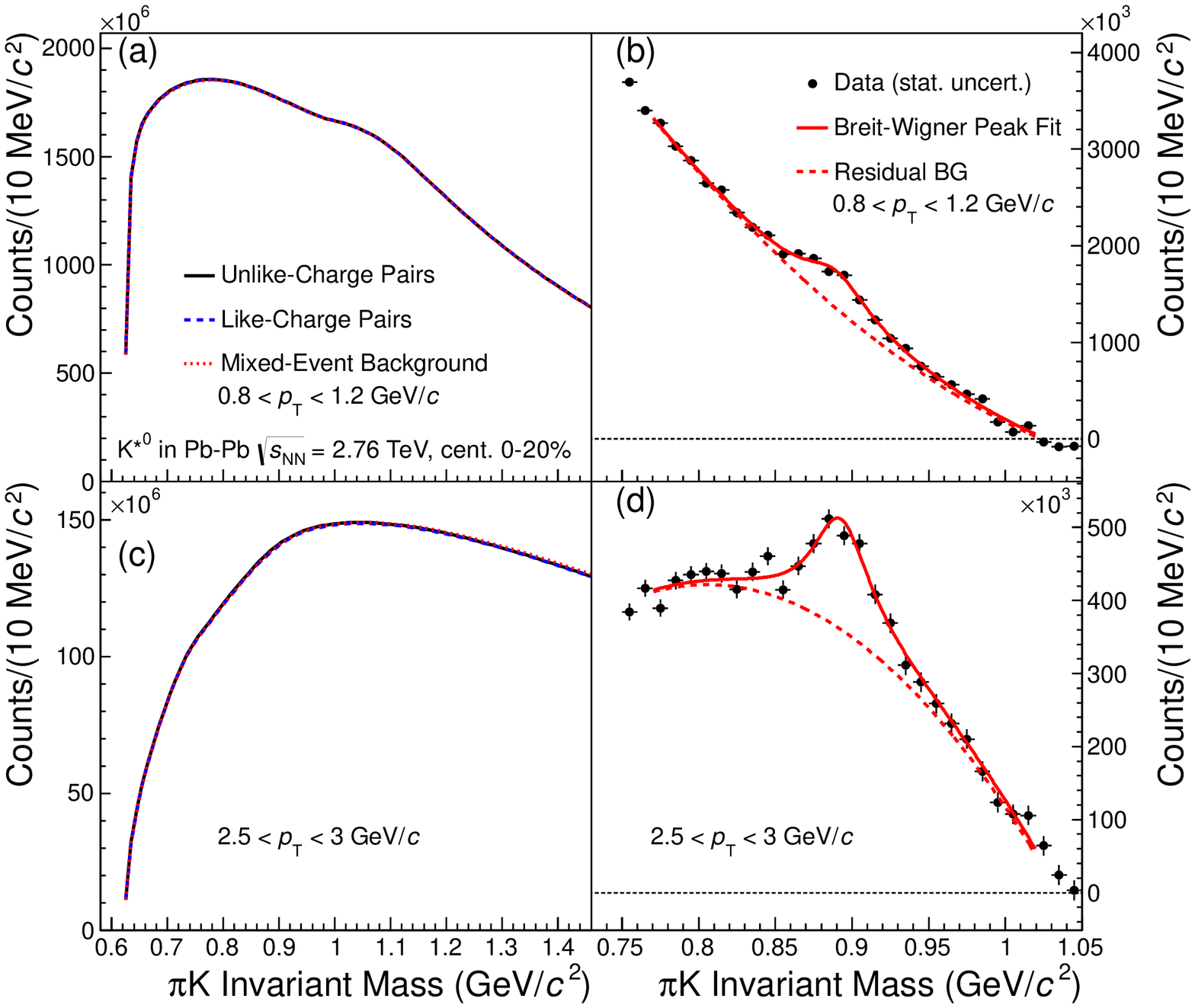}
\caption{\label{fig:signal:examples_ks}(Color online)  Example invariant-mass distributions for the \ks in the 0-20\% centrality interval in two \ptt ranges: $0.8<\ptt<1.2$~\gvc [panels (a) and (b)] and $2.5<\ptt<3$~\gvc [panels (c) and (d)].  Panels (a) and (c) show the unlike-charge invariant-mass distributions for \ks with combinatorial backgrounds.  The normalized mixed-event combinatorial background is within 0.5\% (0.7\%) of the unlike-charge distribution for the low (high) \ptt bin over the invariant-mass range shown here.  The statistical uncertainties are not visible given the vertical scale.  Panels (b) and (d) show the invariant-mass distributions after subtraction of the mixed-event background (plotted with statistical uncertainties) with fits to describe the peaks of the \ks (solid curves) and residual backgrounds (dashed curves).  In the interval $0.8<\ptt<1.2$~\gvc ($2.5<\ptt<3$~\gvc), the uncorrected \ks yield is 7.4 (2.4) million, or 2.4 (0.80) per event; the signal-to-background ratio is $1.1\tms 10^{-4}$ ($5.6\tms 10^{-4}$) and the significance of the \ks peak is 17 (25).}
\end{figure*}

\begin{figure*}
\includegraphics[width=38pc]{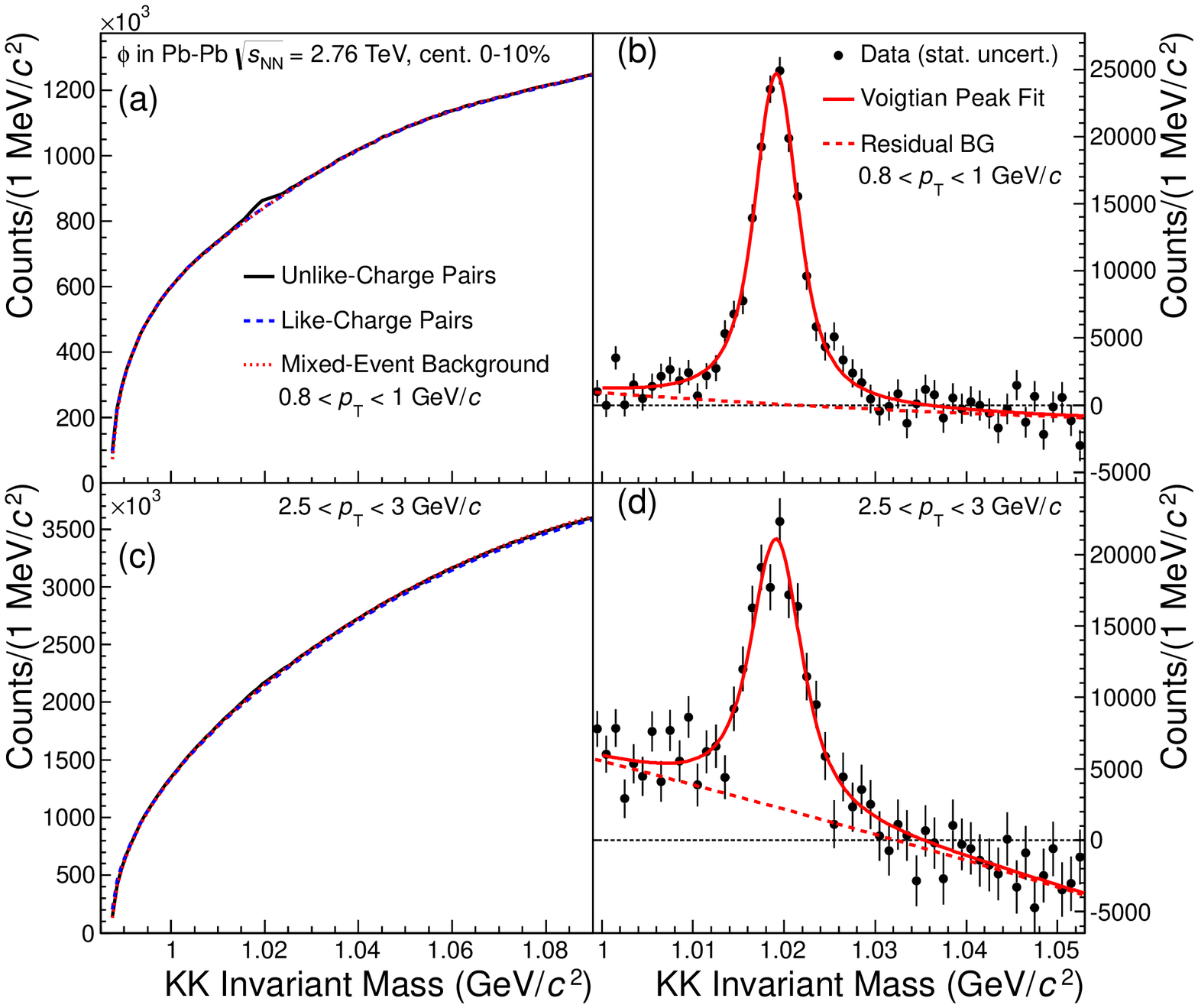}
\caption{\label{fig:signal:examples_phi}(Color online) Example invariant-mass distributions for the \ph in the 0-10\% centrality interval in two \ptt ranges: $0.8<\ptt<1$~\gvc [panels (a) and (b)] and $2.5<\ptt<3$~\gvc [panels (c) and (d)].  Panels (a) and (c) show the unlike-charge invariant-mass distributions for \ph with combinatorial backgrounds.  The normalized mixed-event combinatorial background is within 0.5\% (1\%) of the unlike-charge distribution for the low (high) \ptt bin over most of the invariant-mass range shown here (with the exception of the \ph peak itself and $\mkk<$~0.995~\gvcc).  The statistical uncertainties are not visible given the vertical scale.  Panels (b) and (d) show the invariant-mass distributions after subtraction of the mixed-event background (plotted with statistical uncertainties) with fits to describe the peaks of the \ph (solid curves) and residual backgrounds (dashed curves).  In the interval 0.8~$<\ptt<$~1~\gvc ($2.5<\ptt<3$~\gvc), the uncorrected \ph yield is 174,000 (149,000), or 0.11 (0.095) per event; the signal-to-background ratio is 0.01 (0.0035) and the significance of the \ph peak is 38 (21).}
\end{figure*}

The \ks and \ph resonances are reconstructed through their invariant mass via identified decay-product candidates.  For each centrality and \ptt interval, the invariant-mass distribution of pairs of unlike-charge resonance decay products from the same event is constructed [as an example, see Figs.~\ref{fig:signal:examples_ks} and \ref{fig:signal:examples_phi}, panels (a) and (c)].  It is required that the rapidity of the pair lies within the range $|y_{\mathrm{pair}}|<0.5$.  In the construction of the \ks invariant-mass distributions, it is possible that a track will be designated as both a pion candidate and a kaon candidate because it passes both identification cuts (especially at high \ptt).  In this event, such a track is assigned the kaon mass for some pairs and the pion mass for other pairs.  First, the track will be assigned the kaon mass and pairs will be formed with each of the pion candidate tracks.  Then the track will be assigned the pion mass and paired with each of the kaon candidates.  (The track will never be paired with itself.)  The \ks (\ph) peak has a signal-to-background ratio\footnote{The signal-to-background ratio is evaluated by comparing the integrals of the signal and background over the ranges $0.77<\mpik<1.02$~\gvcc for the \ks and $1.01<\mkk<1.03$~\gvcc for the \ph.} that ranges from $1.1\tms 10^{-4}$ $(1.4\tms 10^{-3})$ to 0.049 (1.7), depending on the \ptt interval analyzed.  For the full \ptt range and centrality 0-80\%, the \ks (\ph) peak has a signal-to-background ratio of $2.7\tms 10^{-4}$ ($4.4\tms 10^{-3}$).

The combinatorial background is estimated with an event-mixing technique by forming pairs using particles from different events.  Each decay-product candidate track is combined with tracks from five other events to build uncorrelated pairs.  Events for mixing are grouped based on the following similarity criteria: the difference in the vertex $z$ position is less than 2~(5)~cm for the \ks (\ph) and the difference in the centrality percentile is required to be less than 10\%.  For the \ks analysis, the difference in the event plane azimuthal angles between the two events is required to be less than $30^{\circ}$.  The signal-to-background ratio is lower for the \ks than the \ph and the residual background for the \ks also tends to have a larger slope or greater curvature than for the \ph.  For these reasons, and in order to provide a mixed-event combinatorial background which is a good representation of the true combinatorial background, the event mixing similarity criteria are somewhat stricter for the \ks.  The \ks mixed-event combinatorial background is normalized such that its integral in the region of $1.1<\mpik<1.3$~\gvcc is the same as the integral of the unlike-charge distribution over the same interval. The \ph mixed-event combinatorial background is normalized to a region that surrounds, but excludes, the \ph peak ($1<\mkk<1.01~\gvcc$ and $1.03<\mkk<1.06~\gvcc$).  The boundaries of the normalization regions are changed and the resulting variations in the experimental results (\textit{e.g.}, average values of 2.2\% for the \ks yield and 0.4\% for the \ph yield) are incorporated into the systematic uncertainties (see ``Combinatorial background" in Table~\ref{table:sys}).  The combinatorial background is also estimated from the invariant-mass distribution of like-charge pairs from the same event.  However, the resulting yields have larger statistical uncertainties and larger bin-to-bin fluctuations than the mixed-event background; the latter is therefore used for this analysis.  Due to its lower signal-to-background ratio, the analysis of the \ks is performed in four centrality intervals from 0-80\%, while the \ph analysis is performed in narrower centrality intervals.

After the normalized combinatorial background has been subtracted from the unlike-charge distribution, \ks and \ph peaks can be observed on top of a residual background [as an example, see Figs.~\ref{fig:signal:examples_ks} and \ref{fig:signal:examples_phi}, panels (b) and (d)].  The residual background may be due to correlated $\pion\mathrm{K}$ or KK pairs emitted within a jet, correlated pairs from particle decays (with three or more stable particles at the end of the decay chain), or misidentified correlated pairs (\textit{e.g.}, a $\uprho\rightarrow\pion\pion$ decay being misidentified as a $\ks\rightarrow\pion\mathrm{K}$ decay).  Differences in the structure of the two mixed events, including differences in the event planes, elliptic flow, primary vertices, and multiplicities, can also lead to an imperfect combinatorial background (if necessary, such differences can be reduced through the use of similarity criteria for the mixed events as described above).  Figures~\ref{fig:signal:examples_ks} and~\ref{fig:signal:examples_phi} show invariant-mass distributions for the \ks and \ph mesons, respectively (two \ptt intervals each).  Integrated over the full transverse-momentum range and using the same centrality interval of 0-80\% for both particles, the uncorrected \ks (\ph) yield is 27.4 (5.9) million, or 2.2 (0.47) per event, with a significance of 86 (146).  For each \ptt and centrality interval, the background-subtracted invariant-mass distributions are fitted by using a combined function to describe the residual background and the signal peak (the peak fitting functions are described below).  The fitting regions are $0.77<\mpik<1.02$~\gvcc for the \ks and $1<\mkk<1.07~\gvcc$ for the \ph.  The boundaries of the fitting region are varied by 10-50~\mvcc for \ks and 5-30~\mvcc for the \ph.  The variation in the yields does not increase if the fitting region boundaries are varied by larger amounts.  Varying the boundaries of the fitting region produces average variations in the \ks (\ph) yield of 9.9\% (3.5\%), which are added to the systematic uncertainties (``Fitting region" in Table~\ref{table:sys}).  The systematic uncertainties also include variations due to the order of the residual background polynomial (first-\nolinebreak, second-\nolinebreak, or third-order).  Varying the residual background polynomial changes the \ks (\ph) yield by 5.8\% (2.7\%) on average (``Residual background shape" in Table~\ref{table:sys}).

For each \ptt and centrality interval, the \ks mass and width are extracted from a relativistic $p$-wave Breit-Wigner function with a Boltzmann factor.

\begin{equation}
\label{eq:signal:pwavebw}
\frac{\dd N}{\dd m_{\pion\mathrm{K}}}=\frac{Cm_{\pion\mathrm{K}}\Gamma M_{0}}{(m_{\pion\mathrm{K}}^{2}-M_{0}^{2})^{2}+M_{0}^{2}\Gamma^{2}}\left[\frac{m_{\pion\mathrm{K}}}{\sqrt{m_{\pion\mathrm{K}}^{2}+\ptt^{2}}}\exp\left(-\frac{\sqrt{m_{\pion\mathrm{K}}^{2}+\ptt^{2}}}{T}\right)\right].
\end{equation}

Here, $C$ is an overall scale factor and $M_{0}$ is the pole mass.  The Boltzmann factor [in square brackets in Eq.~(\ref{eq:signal:pwavebw})] is based on the assumption that in \ada collisions the \ks resonance is predominantly produced through scattering (\textit{e.g.}, $\pion\mathrm{K}\rightarrow\ks$) in a thermalized medium rather than directly from string fragmentation.  The factor accounts for the phase-space population of the parent pions and kaons~\cite{STAR_Kstar_200GeV_2005,ZXu_resonances_RHIC,Kolb,Shuryak}.  The temperature $T$ is fixed to 160~MeV; this is approximately equal to the chemical freeze-out temperature and varying this temperature by $\pm30$~MeV does not produce a significant change in the \ks mass position.  The parameter $\Gamma$ in Eq.~(\ref{eq:signal:pwavebw}) is not constant, but depends on $m_{\pion\mathrm{K}}$, the pole mass $M_{0}$, the resonance width $\Gamma_{0}$, and the vacuum masses of the charged pion and charged kaon ($M_{\pion}$ and $M_{\mathrm{K}}$, respectively)

\begin{equation}
\label{eq:signal:Gamma}
\Gamma=\Gamma_{0}\frac{M_{0}^{4}}{m_{\pion\mathrm{K}}^{4}}\left[\frac{\left(m_{\pion\mathrm{K}}^{2}-M_{\pion}^{2}-M_{\mathrm{K}}^{2}\right)^{2}-4M_{\pion}^{2}M_{\mathrm{K}}^{2}}{\left(M_{0}^{2}-M_{\pion}^{2}-M_{\mathrm{K}}^{2}\right)^{2}-4M_{\pion}^{2}M_{\mathrm{K}}^{2}}\right]^{3/2}.
\end{equation}

The \ks yield is determined by integrating the background-subtracted invariant-mass distribution over the range $0.77<\mpik<1.02$~\gvcc, removing the integral of the residual background fit over the same range, and correcting the result to account for the yield outside that range.  For this purpose, the \ks peak is fitted with a non-relativistic Breit-Wigner function with the width fixed to the vacuum value, allowing the yield in the tails outside the range of integration to be calculated.  This corresponds to $\sim 9\%$ of the total \ks yield.  As an alternative, the \ks yield is also found by integrating the peak fitting functions.  The systematic uncertainties of the \ptt-differential \ks yield, the \ptt-integrated yield \dndy, and the mean transverse momentum \mpt account for variations due to the two methods applied in extracting the yield.  This variation is 2.5\% for the \ptt-differential \ks yield (``Yield extraction" in Table~\ref{table:sys}).  The \ks yield is also extracted from a relativistic Breit-Wigner function and a non-relativistic Breit-Wigner function with a free width.  Changes in the experimental results due to these different peak fitting functions are incorporated into the systematic uncertainties.  The \ks yield varies by 5.2\% on average when different peak fitting functions are used (``Peak shape" in Table~\ref{table:sys}).

To find the \ph mass and width for each \ptt and centrality interval, the peak is fitted by using a Voigtian function.\footnote{The choice of fitting functions for the two resonances is driven by the different widths.  The \ks has a width much larger than the resolution; therefore, a Voigtian fit is not necessary.  However, since the \ks is broad enough, its shape may be influenced by phase-space effects.  The \ph has a width of the same order of magnitude as the resolution and phase-space effects can be neglected.}  This is the convolution of a non-relativistic Breit-Wigner peak and a Gaussian, which accounts for the detector resolution

\begin{equation}
\label{eq:signal:voigt}
\frac{\dd N}{\dd m_{\mathrm{KK}}}=\frac{C\Gamma_{0}}{(2\pi)^{3/2}\sigma}\!\int\limits_{-\infty}\limits^{\infty}\exp\left[-\frac{\left(m_{\mathrm{KK}}-m^{\prime}\right)^{2}}{2\sigma^{2}}\right]\frac{1}{\left(m^{\prime}-M_{0}\right)^{2}+\Gamma_{0}^{2}/4}\dd m^{\prime}.
\end{equation}

The mass resolution parameter $\sigma$, which has been shown to be independent of collision centrality, has been constrained to the value extracted from fits of simulated \ph signal peaks.  This value is about 1.2~\mvcc for $\ptt\approx 0.6$~\gvc.  It reaches a minimum of about 1~\mvcc for $\ptt\approx 1.2$~\gvc and increases to about 1.5~\mvcc for $\ptt=4-5$~\gvc.  To estimate $\sigma$, the production and decay of \ph mesons are simulated using HIJING~\cite{HIJING}, while the propagation of the decay products through the \linebreak ALICE detector material is described using GEANT~3~\cite{GEANT3}.  The \ph yield is determined through the same procedure used for the \ks.  The range of integration is $1.01<\mkk<1.03$~\gvcc.  The yield in the tails is about $\sim 13\%$ of the total \ph yield, which is computed using the same Voigtian fits that are used to find the mass and width.  Average variations in the \ph yield of 1.2\% are observed for the two different yield extraction methods.  Different peak shapes are used in order to obtain alternate measurements of the yield, mass, and width.  The resolution $\sigma$ is varied within the range of values observed in the simulation.  Fits are also performed with the width fixed to the vacuum value while the resolution is kept as a free parameter.  On average, the \ph yield varies by 3.3\% when different peak fitting functions are used.

\renewcommand{\arraystretch}{1.1}

\setlength{\tabcolsep}{6pt}
\begin{table*}
\begin{tabular}{ r r r r r r r r r r r }
\hline\hline
 & \multicolumn{2}{ c }{\ddn} & \multicolumn{2}{ c }{\dndy} & \multicolumn{2}{ c }{\mpt} & \multicolumn{2}{ c }{Mass}  & \multicolumn{2}{ c }{Width}\\
Type & \multicolumn{1}{c}{\ks} & $\ph\;$ & \multicolumn{1}{c}{\ks} & $\ph\;$ & \multicolumn{1}{c}{\ks} & $\ph\;$ & \multicolumn{1}{c}{\ks} & $\ph\;\;\;\;$ & \multicolumn{1}{c}{\ks} & $\ph\;$ \\\hline
Combinatorial background & 2.2 & 0.4 & 1.0 & 0.4 & 0.01 & 0.3 & 0.1 & 0.0001 & 2.0 & 0.4 \\
Fitting region & 9.9 & 3.5 & 6.2 & 2.7 & 5.7 & 0.9 & 0.4 & 0.0023 & 18.2 & 4.4 \\
Residual background shape & 5.8 & 2.7 & 2.1 & 1.2 & 2.2 & 0.7 & 0.3 & 0.0025 & 15.7 & 3.9 \\
Yield extraction & 2.5 & 1.2 & 1.4 & 0.6 & 1.2 & 0.3 & -- & -- & -- & -- \\
Peak shape & 5.2 & 3.3 & 2.7 & 2.7 & 1.8 & 0.5 & 0.3 & 0.0007 & 10.0 & 7.8 \\
Particle identification & 2.7 & 6.2 & 1.2 & 2.3 & 1.1 & 2.1 & 0.3 & 0.0130 & 4.4 & 10.3 \\
Tracking/track selection & 10.0 & 10.0 & 10.0 & 10.0 & -- & -- & 0.4 & 0.0038 & 3.0 & 5.5 \\
Material budget & 1.0 & 1.0 & 1.0 & 1.0 & -- & -- & 0.2 & 0.0100 & -- & -- \\
\ptt extrapolation & -- & -- & 1.2 & 2.1 & 2.1 & 1.2 & -- & -- & -- & -- \\
Normalization & 2.7 & 3.3 & 2.7 & 3.3 & -- & -- & -- & -- & -- & -- \\
Total & 17.3 & 14.0 & 12.2 & 11.9 & 7.2 & 3.1 & 0.7 & 0.0192 & 26.4 & 16.7 \\\hline\hline
\end{tabular}
\caption{Values of systematic uncertainties (\%) averaged over all \ptt and centrality intervals for the yield $[\ddn]$, mass, and width and averaged over all centrality intervals for \dndy and \mpt.  The descriptions in the first column give the type of systematic uncertainty (see also the discussion in the text).  \textbf{Combinatorial background:} mixed-event normalization region (Sec.~\ref{sec:signal}). \textbf{Fitting region:} region used to fit invariant-mass peaks (Sec.~\ref{sec:signal}). \textbf{Residual background shape:} residual background fitting function (Sec.~\ref{sec:signal}). \textbf{Yield extraction:} resonance yield extraction method (Sec.~\ref{sec:signal}). \textbf{Peak shape:} see Sec.~\ref{sec:signal}. \textbf{Particle identification:} \dedx cuts to identify decay products (Sec.~\ref{sec:corr}). \textbf{Tracking/track selection:} see Sec.~\ref{sec:corr}. \textbf{Material budget:} see Sec.~\ref{sec:corr}. \textbf{$\pmb{\ptt}$ extrapolation:} \ptt distribution fitting function used for extrapolation (Sec.~\ref{sec:results:spectra}). \textbf{Normalization:} see Sec.~\ref{sec:results:spectra}.  The \ph mass includes an additional 0.01\% systematic uncertainty coming from the uncertainty in the simulated \ph mass (Sec.~\ref{sec:results:mass_width}).  \textbf{``Total"} gives the average over all centrality and \ptt intervals of the total systematic uncertainty.  A dash (--) indicates that a particular type of uncertainty is not relevant for the given quantity.}
\label{table:sys}
\end{table*}

\section{Yield Corrections\label{sec:corr}}

\begin{figure*}
\includegraphics[width=38pc]{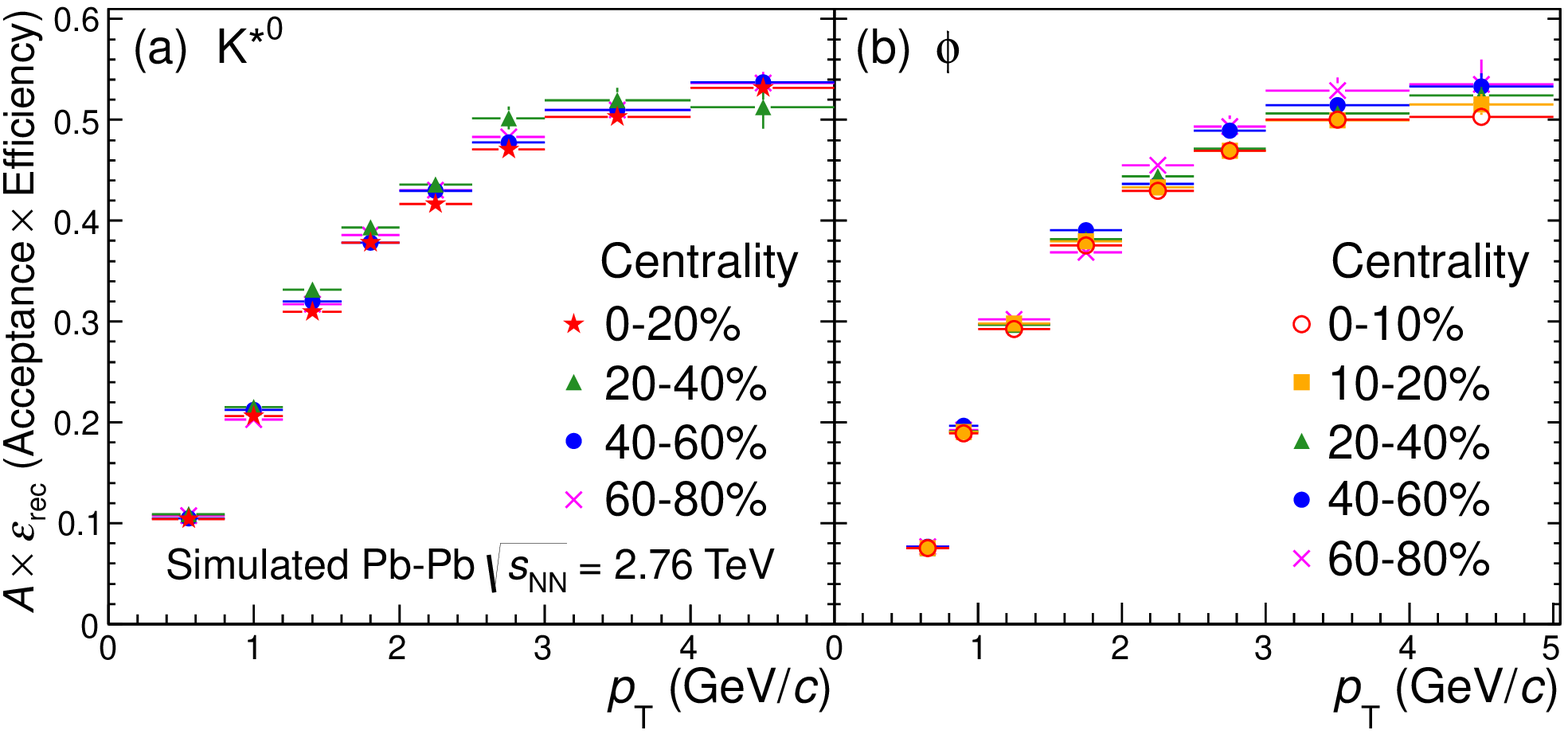}
\caption{\label{fig:corr:efficiency}(Color online) The product of the acceptance and the resonance reconstruction efficiency $A\times\effr$ as a function of \ptt for \ks (a) and \ph (b) mesons, calculated from simulated \pb collisions at \rsnn.  For the \ph meson, only five examples for wide centrality intervals are shown.  The acceptance $A$ includes the effect of the resonance pair rapidity cut ($|y|<0.5$).  The values shown here do not include the branching ratios.  Only statistical uncertainties are shown.}
\end{figure*}

To obtain the corrected resonance yields, the raw yields are divided by the decay branching ratios~\cite{PDG}, the acceptance $A$, the resonance reconstruction efficiency \effr, and the particle identification (PID) efficiency \effp.  The acceptance accounts for the geometrical acceptance of the ALICE detector, the $|y|<0.5$ resonance rapidity cut, and in-flight decays of the pions and kaons used to reconstruct the resonances.  The PID efficiency accounts for the particle identification cuts used to identify the species of the decay-product candidates, \textit{i.e.}, the \dedx cuts in the TPC.  The factor \effr accounts for the remainder of the efficiency, including the tracking efficiency and the cuts used to select good-quality tracks coming from the primary vertex.  The product $A\times\effr$ is extracted from the same HIJING simulations that are used to estimate the mass resolution (with $9\tms 10^{5}$ generated \ks and $4\tms 10^{5}$ generated \ph mesons).  The factor $A\times\effr$ is the fraction of simulated resonances for which both decay products are reconstructed in the ALICE detector and pass the track selection cuts (PID cuts excluded).  Figure~\ref{fig:corr:efficiency} shows $A\times\effr$ for \ks and \ph mesons as a function of \ptt in different centrality intervals.  The efficiency \effp is the product of the independent \dedx-cut efficiencies for each decay product.  The \dedx distributions of the decay-product candidates are Gaussians with resolution \stpc.  When PID cuts of $2\stpc$ are applied to the \dedx values of the pion and kaon candidates (\textit{i.e.}, for both resonance decay products) $\effp=91.1\%$.  The use of different \dedx cuts (1.5\stpc and 2.5\stpc) can result in large changes in the shape of the residual background, which affects the extracted resonance signal.  The \ks (\ph) yield varies by 2.7\% (6.2\%) on average and these variations are incorporated into the systematic uncertainties (see ``Particle identification" in Table~\ref{table:sys}).  A systematic uncertainty of 10\% (for all \ptt and centrality intervals), adapted from the analysis described in~\cite{ALICE_charged_highpT_2760GeV}, accounts for variations in the yields due to the tracking efficiency and different choices of track quality cuts (``Tracking/track selection" in Table~\ref{table:sys}).  A systematic uncertainty of 1\% (for all \ptt and centrality intervals), which accounts for the uncertainty in the yield due to the uncertainty in the material budget  of the ALICE detector (``Material budget" in Table~\ref{table:sys}), is estimated based on~\cite{ALICE_strange_900GeV}.  The uncertainties in the branching ratios~\cite{PDG} are negligible in comparison to the total systematic uncertainties.  The yields extracted with different cuts on the primary vertex $z$ position are found to be consistent with each other.

\section{Results and Discussion\label{sec:results}}

\subsection{Transverse-Momentum Distributions\label{sec:results:spectra}}

\begin{figure*}
\includegraphics[width=35pc]{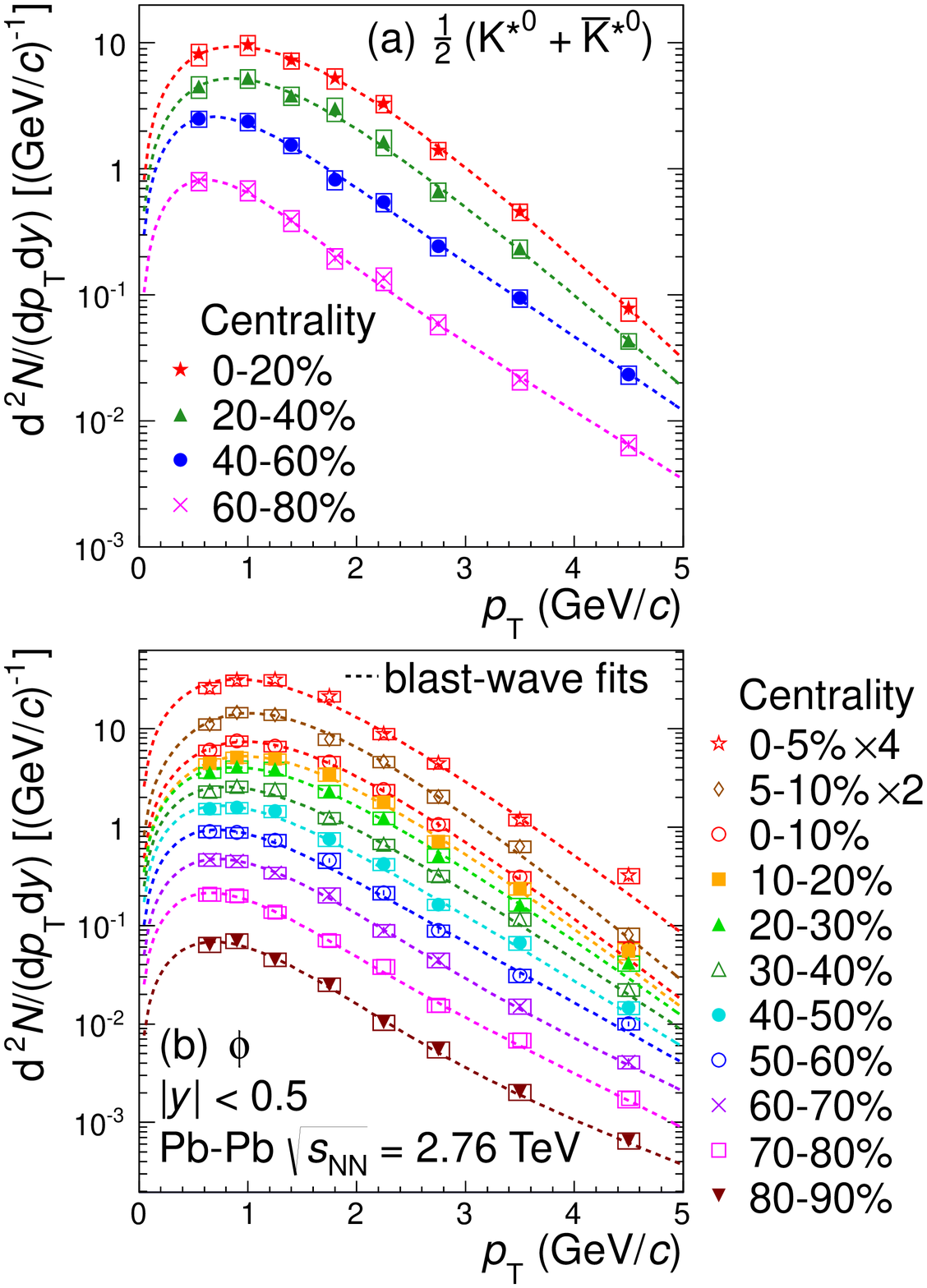}
\caption{\label{fig:spectra}(Color online)  Transverse-momentum distributions of \ks (a) and \ph (b) mesons in multiple centrality intervals with blast-wave fitting functions.  The data are the bin-averaged yields plotted at the bin centers.  The statistical uncertainties are shown as bars and are frequently smaller than the symbol size.  The total systematic uncertainties (including \ptt-uncorrelated and \ptt-correlated components) are shown as boxes.  }
\end{figure*}

The \ptt distributions of the \ks and \ph mesons for $|y|<0.5$, normalized to the number of events and corrected for the efficiency, acceptance, and branching ratio of the decay channel, are shown in Fig.~\ref{fig:spectra}.  For central (peripheral) collisions the statistical uncertainty is approximately 3\% (4\%) near the maximum of the \ptt distribution and increases to approximately 7\% (10\%) in the highest \ptt bin; the systematic uncertainties are summarized in Table~\ref{table:sys}.  In order to extract the values of the mean transverse momentum \mpt and the \ptt-integrated particle yield \dndy, these \ptt distributions are fitted with a Boltzmann-Gibbs blast-wave function~\cite{BoltzmannGibbsBlastWave}, which assumes that the emitted particles are locally thermalized in a uniform-density source at a kinetic freeze-out temperature $T_{\mathrm{kin}}$ and move with a common collective transverse radial flow velocity field.  In this parametrization,

\begin{equation}
\label{eq:results:spectra:blastwave}
\frac{1}{\ptt}\frac{\dd N}{\dd\ptt}\propto \int_{0}^{R}r\;\dd r\;\mT\; I_{0}\negthickspace\left(\frac{\ptt\;\mathrm{sinh}\rho}{T_{\mathrm{kin}}}\right)K_{1}\negthickspace\left(\frac{\mT\;\mathrm{cosh}\rho}{T_{\mathrm{kin}}}\right).
\end{equation}

Here, the transverse mass $\mT=\sqrt{m^{2}+\ptt^{2}}$, $I_{0}$ and $K_{1}$ are modified Bessel functions, $R$ is the fireball radius, and $r$ is the radial distance in the transverse plane.  The velocity profile $\rho$ is

\begin{equation}
\label{eq:results:spectra:rho}
\rho=\mathrm{tanh}^{-1}\beta_{\mathrm{T}}=\mathrm{tanh}^{-1}\left[\left(\frac{r}{R}\right)^{n}\beta_{\mathrm{s}}\right],
\end{equation}

\noindent where $\beta_{\mathrm{T}}$ is the average transverse expansion velocity and $\beta_{\mathrm{s}}$ is the transverse expansion velocity at the surface.  The free parameters in the fits are $T_{\mathrm{kin}}$, $\beta_{\mathrm{s}}$, and the velocity profile exponent $n$.  These fits have $\cn<1.3$ for all centrality intervals.  Between central and peripheral collisions, it is observed that the temperature and the velocity profile exponent $n$ increase, while the expansion velocity decreases, trends which are also observed in blast-wave fits of \pix, \kx, and (anti)proton \ptt distributions in the same collision system~\cite{ALICE_piKp_PbPb}.  The behavior of $T_{\mathrm{kin}}$ and $\beta_{\mathrm{s}}$ as a function of centrality is also observed at RHIC~\cite{STAR_overview_2005,PHENIX_phi_AuAu_2005}.  These trends are consistent with a scenario in which the fireballs created in peripheral collisions have shorter lifetimes than in central collisions, with higher freeze-out temperatures and less time to build up radial flow~\cite{Heinz_Lecture2003}.

In order to find \dndy the measured resonance \ptt distributions are integrated, while the fits are \linebreak used to estimate the resonance yields at low and high \ptt, where no signal could be measured.  The low-\ptt extrapolation region [$\ptt(\ks)<0.3$~\gvc and $\ptt(\ph)<0.5$~\gvc] accounts for 5\% (14\%) of the total \ks (\ph) yield, while the high-\ptt extrapolation region ($\ptt>5$~\gvc) accounts for $\sim 0.1\%$ $(<0.5\%)$ of the total yield.  Alternate functions are also used to fit the resonance \ptt distributions: \linebreak L\'{e}vy-Tsallis functions~\cite{Tsallis,STAR_strange_pp_2007} for both resonances as well as exponential functions in transverse mass for the calculation of \dndy for \ph.  Variations in \dndy and \mpt due to the choice of the fitting function are incorporated into the systematic uncertainties (``\ptt extrapolation" in Table~\ref{table:sys}).  The values of \dndy for \ks (\ph) vary by 1.2\% (2.1\%) on average when the alternate fitting functions are used.  Uncertainties in the boundaries of the centrality percentiles result in a normalization uncertainty for the particle yields.  The values of the normalization uncertainty reported in~\cite{ALICE_piKp_PbPb} (ranging from 0.5\% for central collisions to $_{-8.5}^{+12}\%$ for peripheral collisions) are also used for \ks and \ph (``Normalization" in Table~\ref{table:sys}).

\subsection{Mass and Width\label{sec:results:mass_width}}

\begin{figure*}[h!]
\includegraphics[width=38pc]{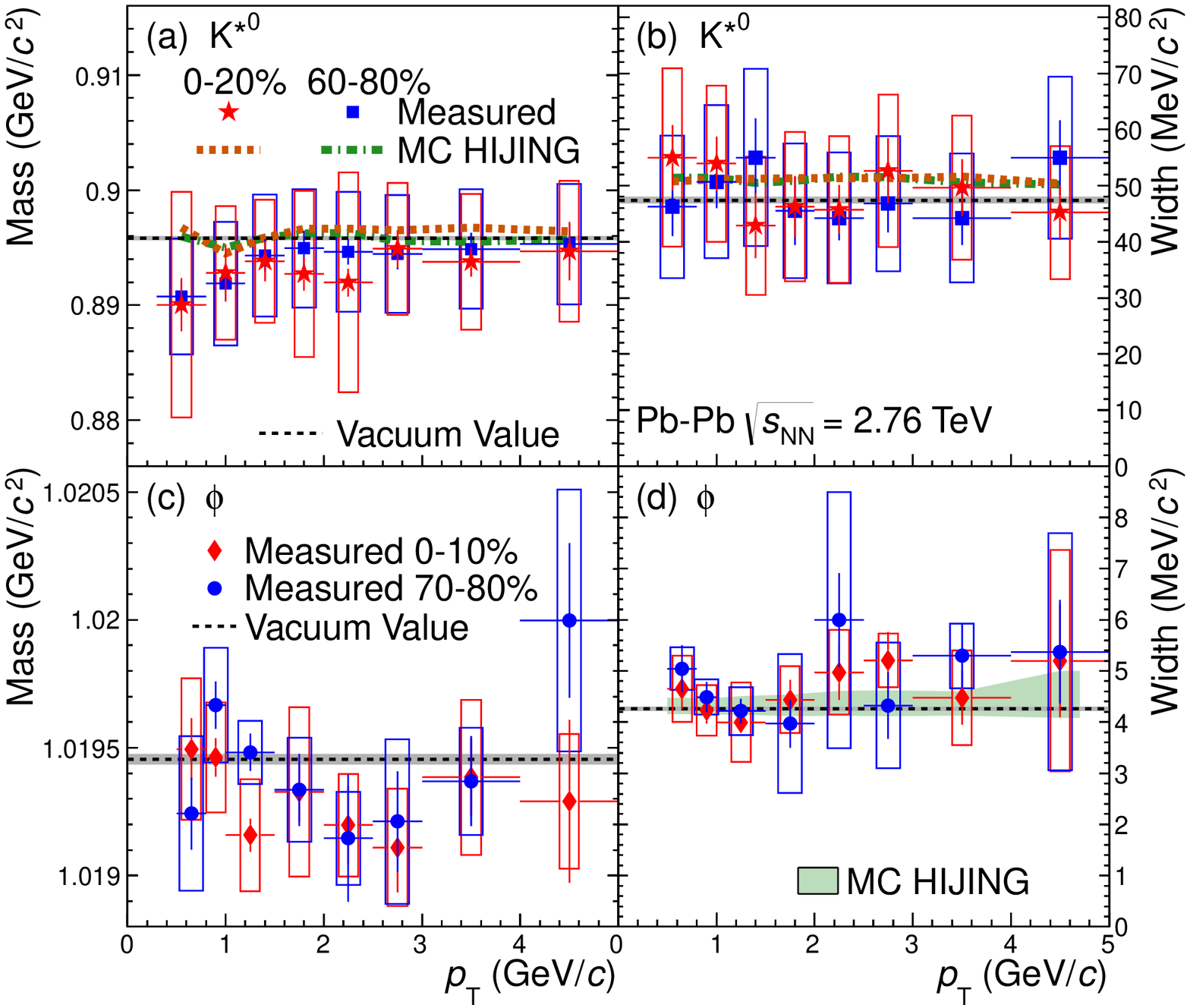}
\caption{\label{fig:results:mass_width}(Color online)  Measured \ks meson mass (a) and width (b) in \pb collisions at \rsnn in the 0-20\% and 60-80\% centrality intervals, along with the values extracted from Monte-Carlo HIJING simulations.  Measured \ph meson mass (c) and width (d) in \pb collisions at \rsnn in the 0-10\% and 70-80\% centrality intervals.  The \ph width extracted from HIJING simulations is also shown.  The vacuum values of the \ks and \ph mass and width~\cite{PDG} are indicated by the horizontal dashed lines.  The statistical uncertainties are shown as bars and the total systematic uncertainties (including \ptt-uncorrelated and \ptt-correlated components) are shown as boxes.}
\end{figure*}

The masses and widths of the \ks and \ph resonances [\textit{i.e.}, the fit parameters $M_{0}$ and $\Gamma_{0}$ from Eqs.~(\ref{eq:signal:pwavebw})-(\ref{eq:signal:voigt})] are shown in Fig.~\ref{fig:results:mass_width} as a function of \ptt for multiple centrality intervals.  The systematic uncertainties in the masses and widths are evaluated as described in Secs.~\ref{sec:signal} and~\ref{sec:corr}.  In addition, variations in the masses and widths of the resonances due to changes in the track selection cuts (on $\mathrm{DCA}_{xy}$ and the number of TPC clusters) are incorporated into the systematic uncertainties (average values of 0.4\% for the \ks mass and 0.0038\% for the \ph mass).  Uncertainties in the material budget of the ALICE detector introduce a further systematic uncertainty of approximately 0.2\% (0.01\%) in the \ks (\ph) mass.  The measured \ks mass has uncertainties of 5-10~\mvcc (an uncertainty of approximately 4~\mvcc is correlated between \ptt bins) and is consistent with the mass values found in the HIJING simulation.  The measured \ks width has uncertainties of 10-20~\mvcc (2~\mvcc correlated between \ptt bins) and is also consistent with the values found in the simulation.  The width of the \ph meson is an order of magnitude smaller than the width of the \ks.  The \ph mass is therefore measured with better precision than the \ks, with systematic uncertainties of $\sim0.2$~\mvcc.  A mass shift, due to detector effects, is observed in the HIJING simulation.  This shift ranges from $-0.35$~\mvcc at low \ptt to +0.05~\mvcc at high \ptt.  The measured \ph mass is corrected to account for this shift.  The corrected \ph mass, shown in Fig.~\ref{fig:results:mass_width}(c), has uncertainties of 0.15-0.5~\mvcc  (0.1~\mvcc correlated between \ptt bins).  The \ph mass is observed to be consistent with the vacuum value.  The \ph width has uncertainties of 0.7-2~\mvcc (0.3~\mvcc correlated between \ptt bins) and is consistent with the width observed in the HIJING simulation.\footnote{No centrality dependence is observed for the \ph width in the simulation, so the average width for centrality 0-80\% is plotted in Fig.~\ref{fig:results:mass_width}(d).}  Neither the mass nor the width of either resonance varies with centrality and no evidence is seen for a modification of the mass or width in \pb collisions at \rsnn.  The masses and widths of these resonances have also been studied at lower collision energies.  No significant change in the mass or width of the \ks meson is observed by the STAR Collaboration in \au and \cu collisions at \rsnn[62.4~GeV] and \rsnn[200~GeV]~\cite{STAR_Kstar_2011}.  The STAR Collaboration observes that the measured mass and width of the \ph meson deviate from the values extracted from simulations at low \ptt ($\lesssim1.5$~\gvc) in pp, \dau, and \au collisions at \rsnn[200~GeV] and \au collisions at \rsnn[62.4~GeV]~\cite{STAR_phi_2009}.  However, the deviations do not appear to depend on the size of the collision system and are likely due to detector effects that are not properly reproduced in the simulations.  No clear evidence is observed for changes in the \ph mass or width by the PHENIX Collaboration in \au collisions at \rsnn[200~GeV]~\cite{PHENIX_phi_AuAu_2005}, nor by the NA49 Collaboration in \pb collisions at \rsnn[6-17~GeV]~\cite{NA49_phi_2008}.

\renewcommand{\arraystretch}{1.2}

\setlength{\tabcolsep}{9pt}
\begin{table*}
\begin{tabular}{ r c c c }
\hline\hline
\multicolumn{4}{ c }{\ks}\\
Centrality & \dndy & \kskm & \mpt (\gvc) \\\hline
0-20\% & $16.6 \pm 0.6 \pm 2.5 \pm 0.1$ & $0.20 \pm 0.01 \pm 0.03$ & $1.31 \pm 0.04 \pm 0.11$ \\
20-40\% & $9.0 \pm 0.8 \pm 1.1 \pm 0.1$ & $0.24 \pm 0.02 \pm 0.03$ & $1.29 \pm 0.04 \pm 0.11$ \\
40-60\% & $3.9 \pm 0.3 \pm 0.4 \pm 0.1$ & $0.28 \pm 0.02 \pm 0.03$ & $1.16 \pm 0.04 \pm 0.08$ \\
60-80\% & $1.13 \pm 0.09 \pm 0.11 \pm0.07$ & $0.31 \pm 0.02 \pm 0.03$ & $1.08 \pm 0.03 \pm 0.07$ \\\hline
\multicolumn{4}{ c }{}\\
\multicolumn{4}{ c }{\ph}\\
Centrality & \dndy & \phikm & \mpt (\gvc) \\\hline
0-5\% & $13.8 \pm 0.5 \pm 1.7 \pm 0.1$ & $0.127 \pm 0.004 \pm 0.014$ & $1.31 \pm 0.04 \pm 0.06$ \\
5-10\% & $11.7 \pm 0.4 \pm 1.4 \pm 0.1$ & $0.130 \pm 0.004 \pm 0.014$ & $1.34 \pm 0.04 \pm 0.06$ \\
10-20\% & $9.0 \pm 0.2 \pm 1.0 \pm 0.1$ & $0.134 \pm 0.003 \pm 0.013$ & $1.34 \pm 0.03 \pm 0.04$ \\
20-30\% & $7.0 \pm 0.1 \pm 0.8 \pm 0.1$ & $0.152 \pm 0.003 \pm 0.015$ & $1.29 \pm 0.02 \pm 0.03$ \\
30-40\% & $4.28 \pm 0.09 \pm 0.48 \pm 0.09$ & $0.144 \pm 0.003 \pm 0.014$ & $1.25 \pm 0.03 \pm 0.03$ \\
40-50\% & $2.67 \pm 0.05 \pm 0.30 \pm 0.06$ & $0.148 \pm 0.003 \pm 0.014$ & $1.22 \pm 0.02 \pm 0.05$ \\
50-60\% & $1.49 \pm 0.03 \pm 0.16 \pm 0.05$ & $0.145 \pm 0.003 \pm 0.014$ & $1.20 \pm 0.02 \pm 0.04$ \\
60-70\% & $0.72 \pm 0.02 \pm 0.08 \pm 0.04$ & $0.140 \pm 0.004 \pm 0.013$ & $1.17 \pm 0.03 \pm 0.05$ \\
70-80\% & $0.30 \pm 0.01 \pm 0.04 \pm 0.02$ & $0.133 \pm 0.005 \pm 0.015$ & $1.12 \pm 0.03 \pm 0.03$ \\
80-90\% & $0.097 \pm 0.004 \pm 0.012 _{-0.008}^{+0.012}$ & $0.113 \pm 0.005 \pm 0.014$ & $1.14 \pm 0.05 \pm 0.06$ \\\hline\hline
\end{tabular}
\caption{The values of \dndy, the \kskm and \phikm ratios, and \mpt are presented for different centrality intervals.  In each entry the first uncertainty is statistical.  For \dndy, the second uncertainty is the systematic uncertainty, not including the normalization uncertainty, and the third uncertainty is the normalization uncertainty.  For \kskm, \phikm, and \mpt, the second uncertainty is the total systematic uncertainty.  The ratios are calculated using \km yields from~\cite{ALICE_piKp_PbPb}.}
\label{table:results}
\end{table*}

\begin{figure}
\includegraphics[width=19.1pc]{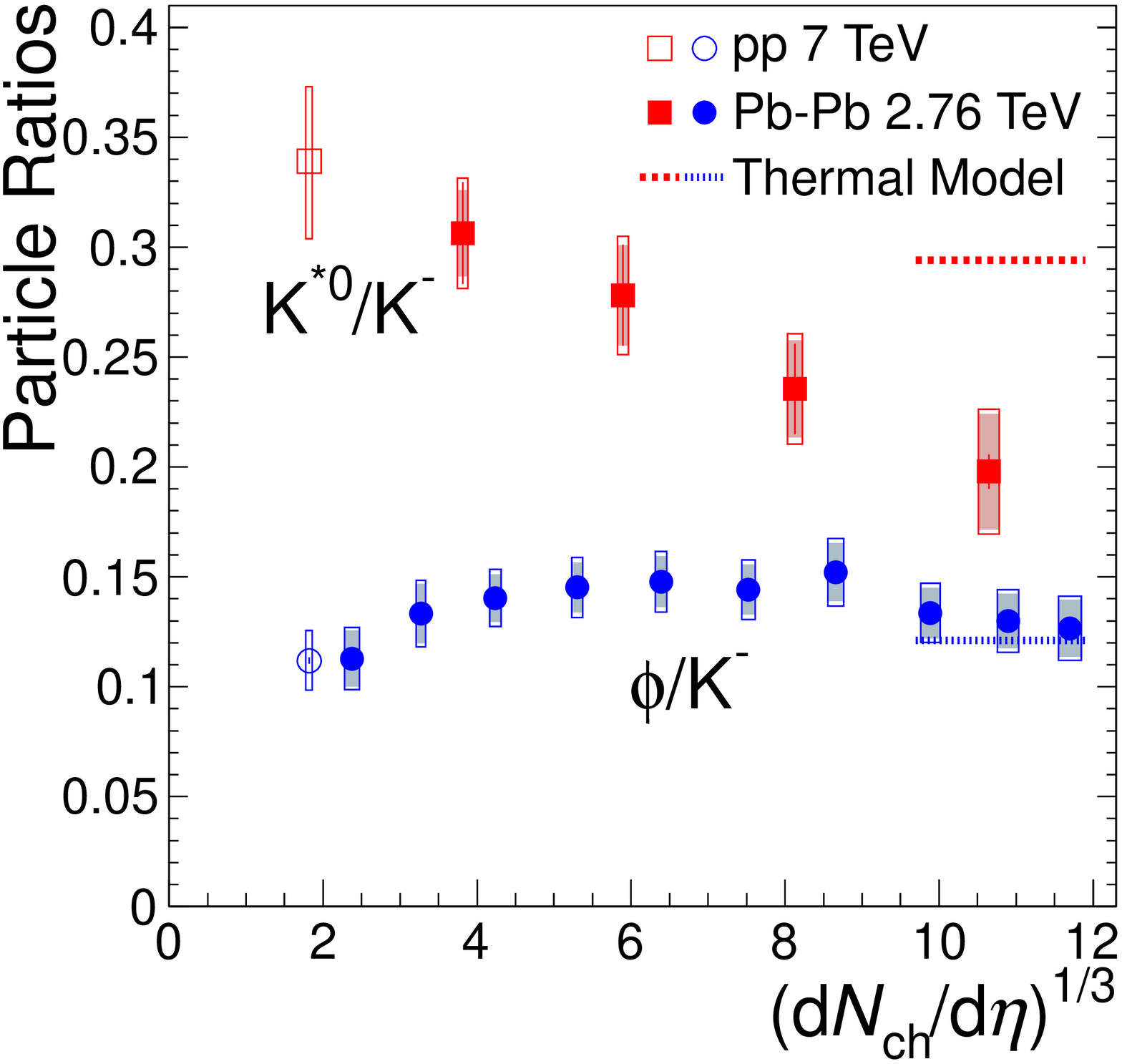}
\hspace{0.5pc}
\begin{minipage}{18pc}
\vspace{-21.65pc}
\caption{\label{fig:results:ratios_cent}(Color online)  Ratios of \ptt-integrated particle yields \kskm and \phikm as a function of \dncr~\cite{ALICE_multiplicity,ALICE_multiplicity_pp7TeV} for \pb collisions at \rsnn and pp collisions at \rs~\cite{ALICE_kstar_phi_7TeV,ALICE_piKp_7TeV}.  The values of \dnc were measured at mid-rapidity.  The statistical uncertainties are shown as bars.  The shaded boxes show systematic uncertainties that are not correlated between centrality intervals, while the open boxes show the total systematic uncertainties including both correlated and uncorrelated sources.  The values given by a grand-canonical thermal model with a chemical freeze-out temperature of 156~MeV are also shown~\cite{Stachel_SQM2013}.}
\end{minipage}
\end{figure}

\subsection{Particle Ratios and Interactions in the Hadronic Phase\label{sec:results:rescattering}}

Table~\ref{table:results} gives the values of \dndy, the \ptt-integrated particle yields for $|y|<0.5$, for the \ks and \ph resonances in different centrality intervals.  This table also includes the ratios of \ptt-integrated particle yields \kskm and \phikm, which are calculated using the \dndy values for \km from~\cite{ALICE_piKp_PbPb}.  These ratios are shown in Fig.~\ref{fig:results:ratios_cent} for \pb collisions at \rsnn and pp collisions at \rs~\cite{ALICE_kstar_phi_7TeV,ALICE_piKp_7TeV}.  These ratios are presented as a function of \dncr (the cube root of the charged-particle multiplicity density measured at mid-rapidity)~\cite{ALICE_multiplicity,ALICE_multiplicity_pp7TeV} for reasons discussed below.  The \kskm ratio is observed to be lower in central \pb collisions [larger values of \dncr] than in pp and peripheral \pb collisions.  When the \kskm ratio in central collisions is divided by the \kskm ratio in peripheral collisions the result\footnote{This calculation excludes the tracking/track selection and material budget systematic uncertainties, which are assumed to be correlated between centrality intervals.} is $0.65\pm0.11$, which is different from unity at the $3.2\sigma$ level.  On the other hand, the \phikm ratio does not depend strongly on collision centrality and may be enhanced in mid-central collisions with respect to peripheral and pp collisions.  The value of the \phikm ratio in central \pb collisions is consistent with the value measured in pp collisions.

As discussed in Sec.~\ref{sec:intro}, it is possible that resonance yields are modified during the hadronic phase by re-scattering (which would reduce the measured yields) and regeneration (which would increase the yields).  The observed suppression of the \kskm ratio may be the result of these effects, with re-scattering dominating over regeneration.  The fact that the \phikm ratio does not exhibit suppression for central collisions suggests that the \ph (which has a lifetime an order of magnitude larger than the \ks) might decay predominantly outside the hadronic medium.  Of \ks mesons with momentum $p=1$~\gvc, 55\% will decay within 5~fm/$c$ of production (a typical estimate for the time between chemical and kinetic freeze-out in heavy-ion collisions~\cite{Bliecher_Aichelin,Bass_Freezeout_1999}), while only 7\% of \ph mesons with $p=1$~\gvc will decay within that time.  It should be noted that elastic scattering of the resonance decay products might be expected to broaden the measured \ks invariant-mass distribution, which is not observed.  The simultaneous observation of \kskm suppression but no \ks width modification could be explained by decay-product re-scattering if that process were to take place predominantly through elastic scattering with large momentum transfers (which would make the modified signal indistinguishable from the background) or through pseudo-elastic scattering via other resonances.

In Fig.~\ref{fig:results:ratios_cent} the \kskm and \phikm ratios have been plotted as a function of \dncr in order to study whether the strength of the suppression might be related to the system radius.  It is an established practice in femtoscopy studies to plot the HBT radii as a function of \dncr~\cite{ALICE_HBT_2011}.  In some cases these radii have been observed to increase approximately linearly with \dncr~\cite{ALICE_HBT_2011,Lisa_FemtoscopyReview}, suggesting that \dncr might be used as a proxy for the system radius.  If it is assumed that the suppression of the \ks yield is due to re-scattering and that the strength of re-scattering effects is proportional to the distance which the decay products travel through the hadronic medium, the \kskm ratio would be expected to decrease as a decaying exponential in \dncr.  The observed dependence of the \kskm ratio on the multiplicity is consistent with the behavior that would be expected if re-scattering were the cause of the suppression.

Figure~\ref{fig:results:ratios_cent} also includes the values given by a thermal model~\cite{Stachel_SQM2013} for the \kskm and \phikm ratios in central \pb collisions at \rsnn, with a chemical freeze-out temperature of 156~MeV and a baryochemical potential of 0~MeV.  This thermal model does not include re-scattering effects.  These results were obtained by fitting a variety of particle yields measured in this collision system.  The \ph yield was included in the fit, but the \ks was excluded due to the possibility that its yield could be modified as discussed above.  The \ph yield from the fit agrees with the measured yield within 0.5 times the uncertainties and the fit results are not expected to change significantly if the \ph is excluded.  The \kskm ratio given by the thermal model is about 50\% larger than the measured ratio.  The thermal-model \phikm ratio for central \pb collisions at \rsnn is consistent with the measured value.

\begin{figure*}[h]
\includegraphics[width=38pc]{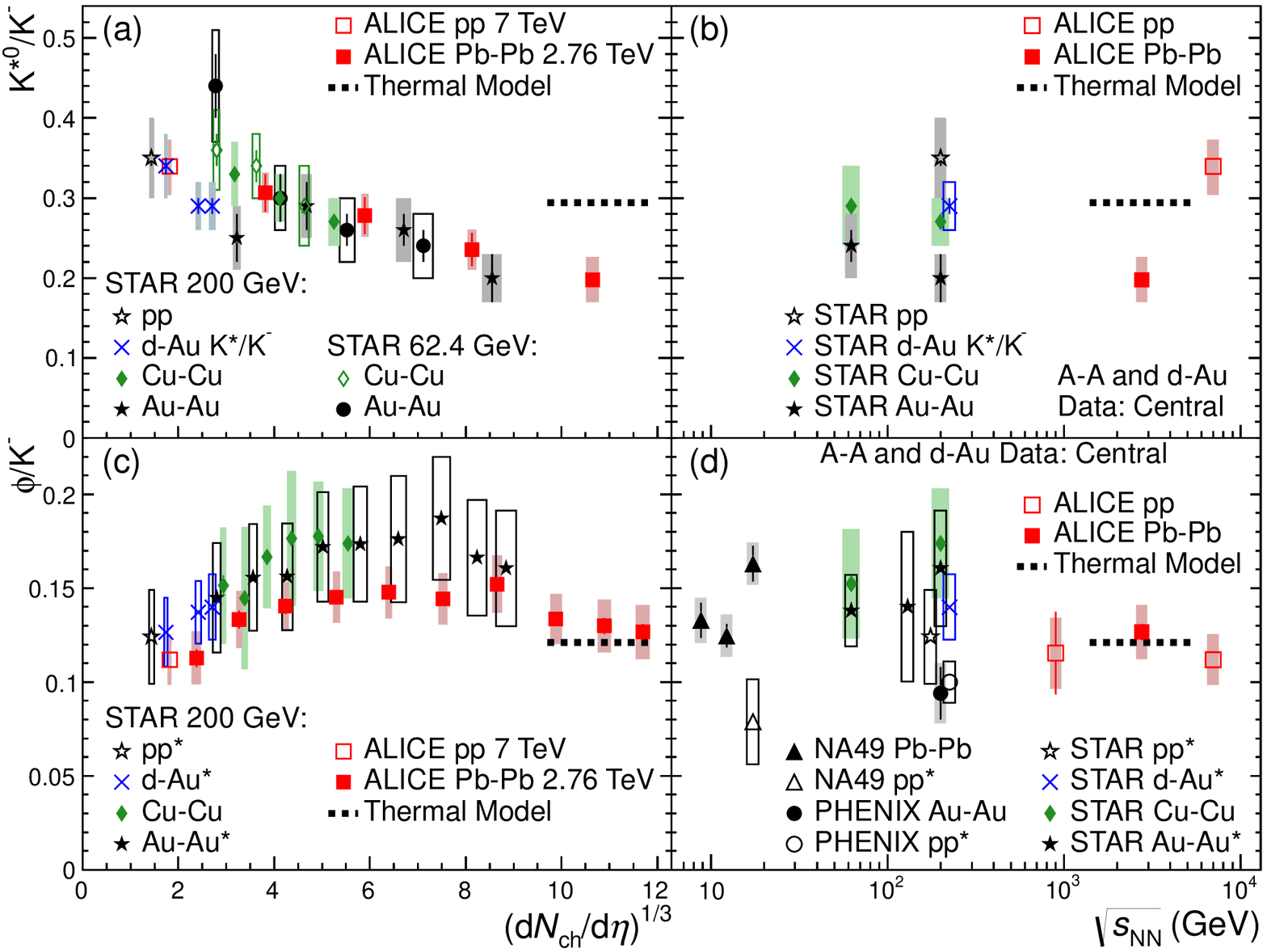}
\caption{\label{fig:results:ratios_full}(Color online)  Particle ratios \kskm [panels (a) and (b)] and \phikm [panels (c) and (d)] in pp, \dau, and \ada collisions~\cite{STAR_Kstar_200GeV_2005,ALICE_strange_900GeV,PHENIX_phi_AuAu_2005,STAR_Kstar_2011,STAR_phi_2009,NA49_phi_2008,ALICE_kstar_phi_7TeV,ALICE_piKp_7TeV,STAR_resonances_dAu_2008,STAR_phi_200GeV_2005,NA49_phi_2000,NA49_piK_2002,STAR_phi_130GeV,PHENIX_mesons_pp_2011,PHENIX_piKp_pp_2011,ALICE_piKp_900GeV}.  In panels (a) and (c) these ratios are presented for different centrality intervals as a function of \dncr~\cite{ALICE_multiplicity,ALICE_multiplicity_pp7TeV,STAR_piKp_CuCu,STAR_centrality_2009}.  The values of \dnc were measured at mid-rapidity.  In panels (b) and (d), these ratios are presented for pp, central \dau, and central \ada collisions as a function of \rsnno.  The values given by a grand-canonical thermal model with a chemical freeze-out temperature of 156~MeV are also shown~\cite{Stachel_SQM2013}.  For quantities marked ``*", boxes represent the total uncertainty (separate uncertainties are not reported).  Otherwise, bars represent the statistical uncertainties and boxes represent the systematic uncertainties (including centrality-uncorrelated and centrality-correlated components).  For the \dau data in panels (a) and (b), the numerator yield is derived from a combination of the charged and neutral $\mathrm{K}^{*}\krr(892)$ states.  In panel (c), the two most central \phikm points for \au collisions are for overlapping centrality intervals (0-5\% and 0-10\%).  The following points have been shifted horizontally for visibility: the lowest-multiplicity \dau points in panels (a) and (c), the \dau points in panels (b) and (d), and the pp data points for \rsnn[200~GeV] in panel (d).}
\end{figure*}

The measured \kskm and \phikm ratios are compared in Fig.~\ref{fig:results:ratios_full} to results for different collision systems and energies, plotted as a function of \dncr and \rsnno.  This figure also includes the same thermal-model ratios for central \pb collisions shown in Fig.~\ref{fig:results:ratios_cent}.  The \kskm ratio is compared in Fig.~\ref{fig:results:ratios_full}(a) and \ref{fig:results:ratios_full}(b) to results for different collision systems at RHIC\footnote{For \dau collisions~\cite{STAR_resonances_dAu_2008} at \rsnn[200~GeV] the ratio $\mathrm{K}^{*}\krl/\mathrm{K}^{-}$ is plotted instead, where the yield in the numerator is calculated from a combination of all four $\mathrm{K}^{*}\krr(892)$ states.}~\cite{STAR_Kstar_200GeV_2005,STAR_Kstar_2011,STAR_resonances_dAu_2008,STAR_piKp_CuCu} and LHC~\cite{ALICE_piKp_PbPb,ALICE_kstar_phi_7TeV,ALICE_piKp_7TeV} energies.  The \kskm ratio is plotted as function of \dncr~\cite{ALICE_multiplicity,ALICE_multiplicity_pp7TeV,STAR_piKp_CuCu,STAR_centrality_2009} in panel (a).  In general, these values appear to follow a single trend independent of collision energy, tending to exhibit suppression in central \ada collisions with respect to pp, \dau, and peripheral \ada collisions.  The decrease in the \kskm ratio between pp and central \ada collisions is similar at both RHIC and LHC energies.  Refs.~\cite{STAR_Kstar_200GeV_2005} and~\cite{STAR_Kstar_2011} also suggest that the decrease in this ratio for collisions at \rsnn[200~GeV] may be due to re-scattering of the \ks decay products in the hadronic medium.  The same ratio is shown in panel (b) as a function of \rsnno for pp collisions, as well as central \ada and \dau collisions.  The \kskm ratio is higher in pp collisions than in central \au and \pb collisions.  The value of the \kskm ratio is larger in central \cu than in central \au collisions, which is expected due to the smaller size of the \cu collision system.

The \phikm ratio is compared in Fig.~\ref{fig:results:ratios_full}(c) and \ref{fig:results:ratios_full}(d) to results for different collision systems at \linebreak SPS~\cite{NA49_phi_2008,NA49_phi_2000,NA49_piK_2002}, RHIC~\cite{PHENIX_phi_AuAu_2005,STAR_phi_2009,STAR_phi_200GeV_2005,STAR_phi_130GeV,PHENIX_mesons_pp_2011,PHENIX_piKp_pp_2011}, and LHC~\cite{ALICE_strange_900GeV,ALICE_kstar_phi_7TeV,ALICE_piKp_7TeV,ALICE_piKp_900GeV} energies.  The \phikm ratio is plotted as a function of \dncr~\cite{ALICE_multiplicity,ALICE_multiplicity_pp7TeV,STAR_piKp_CuCu,STAR_centrality_2009} in panel (c) for collisions at \rsnn[200~GeV] and LHC energies.  The measured \phikm ratio for \ada collisions tends to be larger at \rsnn[200~GeV] than at \rsnn for similar values of \dncr; however, the values are consistent within their uncertainties.  As observed at LHC energies, the \phikm ratio at \rsnn[200~GeV] does not exhibit a strong centrality dependence, though there are indications of a small enhancement (not beyond the uncertainties) for mid-central and central \ada collisions.  The \phikm ratio is shown in panel (d) as a function of \rsnno for pp collisions and for central \ada and \dau collisions.  The \phikm ratio is independent of collision energy and system from RHIC to LHC energies,\footnote{For \au collisions at \rsnn[200~GeV], the \phikm ratio measured by the PHENIX Collaboration~\cite{PHENIX_phi_AuAu_2005} is $\sim40\%$ less than (and not consistent with) the \phikm ratio measured by the STAR Collaboration~\cite{STAR_phi_2009}.  A possible explanation for this discrepancy is discussed in~\cite{Rafelski_2005}.} while at SPS energies the ratio measured in \pb collisions is a factor of two larger than the ratio in pp collisions.

The measured \ptt distributions and yields may reflect elastic and pseudo-elastic interactions in the hadronic phase, with the magnitude of the change depending on the resonance lifetime.  Thermal models, which give particle yields at chemical freeze-out, do not include these effects.  Therefore, including the yields of short-lived resonances like \ks in thermal-model fits might give misleading results.  The model described in~\cite{Markert_thermal,Torrieri_thermal,Torrieri_thermal_2001b,Torrieri_thermal_2001b_erratum} is based on a thermal-model framework, but includes, in addition, re-scattering effects which modify the resonance yields after chemical freeze-out.  This model predicts particle ratios, including \ksk, as a function of the chemical freeze-out temperature and the lifetime of the hadronic phase.  If an assumption is made about the value of the chemical freeze-out temperature, a measured \ksk ratio can be used to extract an estimate of the lifetime.  Assuming a chemical freeze-out temperature of 156~MeV (based on thermal-model fits of ALICE data~\cite{Stachel_SQM2013}) and using the measured \kskm ratio for the 0-20\% centrality interval, it is possible to estimate a lower limit of 2~fm/$c$ for the time between chemical and kinetic freeze-out.  Only a lower limit can be extracted because the model does not include regeneration of resonances in the hadronic medium.  This limit on the hadronic lifetime is the same order of magnitude as the \ks lifetime, but 23 times shorter than the \ph lifetime.  This value can be compared to the hadronic lifetime of $>4$~fm/$c$ extracted using the same model and the $\Lambda(1520)\krl/\krr\Lambda$ ratio measured in \au collisions at \rsnn[200~GeV]~\cite{STAR_baryon_resonances_2006}.  If a constant chemical freeze-out temperature is assumed, the increase in \kskm from central to peripheral \pb collisions (see Fig.~\ref{fig:results:ratios_cent}) corresponds to a decreasing hadronic-phase lifetime and, equivalently, a larger kinetic freeze-out temperature.  This is in qualitative agreement with results from blast-wave fits of particle \ptt distributions~\cite{ALICE_piKp_PbPb}, which also exhibit an increase in the kinetic freeze-out temperature for more peripheral collisions.  Alternatively, if no hadronic lifetime or no re-scattering is assumed, the model predicts a freeze-out temperature for the \ks of about $120\pm7$~MeV.

It should be noted that these estimates of the temperature or the lifetime of the hadronic phase are model-dependent.  The estimate of 2~fm/$c$ for the lower limit of the lifetime of the hadronic phase is only valid insofar as the model described in~\cite{Markert_thermal,Torrieri_thermal,Torrieri_thermal_2001b,Torrieri_thermal_2001b_erratum} is valid.  Later work by one of the same authors~\cite{Petran_Rafelski_2013a,Petran_Rafelski_2013b} uses a non-equilibrium thermal model to extract an estimate of 138~MeV for the freeze-out temperature with no time difference between chemical and kinetic freeze-out.  However, this non-equilibrium model predicts a \ksk ratio that is essentially independent of centrality, which appears to disagree with the results reported above.

\begin{figure*}
\includegraphics[width=38pc]{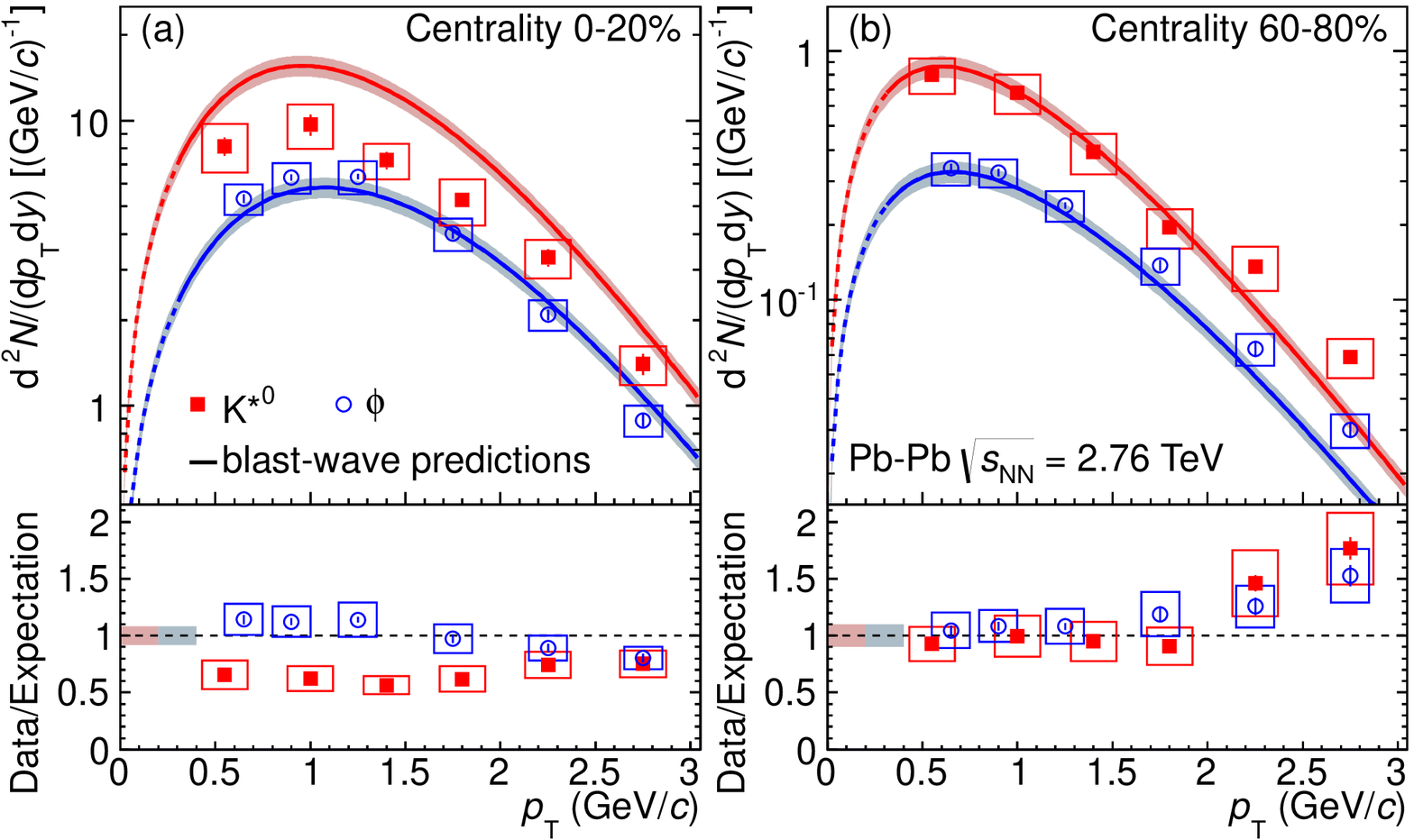}
\caption{\label{fig:results:blast}(Color online)  Transverse-momentum distributions of \ks and \ph resonances in \pb collisions at \rsnn along with expected distributions for central (a) and peripheral (b) collisions.  The shapes of the expected distributions are given by Boltzmann-Gibbs blast-wave functions~\cite{BoltzmannGibbsBlastWave} using parameters obtained from fits to \pix, \kx, and (anti)proton \ptt distributions.  The expected distributions are normalized so that their integrals are equal to the measured yield of charged kaons in \pb collisions~\cite{ALICE_piKp_PbPb} multiplied by the \ksk and \phik ratios given by a thermal-model fit to ALICE data~\cite{Stachel_SQM2013}.  The lower panels show the ratios of the measured distributions to the values from the model.  The statistical uncertainties are shown as bars and the systematic uncertainties from the measured \ptt distributions (including \ptt-uncorrelated and \ptt-correlated components) are shown as boxes.  The shaded bands (upper panels) and shaded boxes (lower panels) indicate the uncertainties in the normalization of the model distributions.}
\end{figure*}

According to UrQMD~\cite{UrQMD_Bass,UrQMD} calculations for RHIC energies, the hadronic re-scattering effect is expected to be momentum-dependent, with greater strength for low-\ptt resonances ($\ptt\lesssim2$~\gvc)~\cite{Bleicher_Stoecker,Bliecher_Aichelin}.  To investigate the \ptt dependence of the observed suppression, the blast-wave model is used to generate expected transverse-momentum distributions without re-scattering effects for the \ks and \ph resonances at kinetic freeze-out.  The kinetic freeze-out temperature $T_{\mathrm{kin}}$, velocity profile exponent $n$, and surface expansion velocity $\beta_{\mathrm{s}}$ (radial flow) are taken from simultaneous blast-wave fits of \pix, \kx, and (anti)proton \ptt distributions in \pb collisions at \rsnn~\cite{ALICE_piKp_PbPb}.  For each centrality interval, these fits were performed over the ranges $0.5<\ptt<1$~\gvc for \pix, $0.2<\ptt<1.5$~\gvc for \kx, and $0.3<\ptt<3$~\gvc for (anti)protons.  The simultaneous fits provide good descriptions of the particle \ptt distributions within these fit ranges.  The parameters used in the present study are the averages (weighted by the number of events multiplied by \dnc~\cite{ALICE_multiplicity}) of the values reported for the narrower centrality intervals in~\cite{ALICE_piKp_PbPb}.  For the 0-20\% (60-80\%) centrality interval, $T_{\mathrm{kin}}$ is 0.097~GeV (0.13~GeV), $n$ is 0.73 (1.38), and $\beta_{\mathrm{s}}$ is 0.88 (0.80).  For $\ptt<3$~\gvc, these parameters are used with the blast-wave model to generate the shapes, but not the total yields, of expected \ptt distributions for the \ks and \ph mesons.  The expected \ks (\ph) distribution is normalized so that its integral is the \kx yield \dndy in \pb collisions~\cite{ALICE_piKp_PbPb} multiplied by the \kskm (\phikm) ratio given by a thermal-model fit to ALICE data~\cite{Stachel_SQM2013} (with a chemical freeze-out temperature of 156~MeV).  These are taken to be the expectations if hydrodynamics, as parameterized by the blast-wave model, describes the \ptt distributions of the stable particles and the resonances simultaneously and if the \ks and \ph meson \ptt distributions are not modified by any additional effects (\textit{e.g.}, re-scattering or regeneration).  The normalization depends on the parameters of the thermal model: if a temperature of 164~MeV~\cite{AndronicQM2011} is used instead, the expected \ks (\ph) yield is 5\% (6\%) greater.

Figure~\ref{fig:results:blast} shows these expected \ks and \ph distributions (as solid lines), the measured resonance \ptt distributions, and the ratios of the measurement to the model for central (0-20\%) and peripheral \linebreak(60-80\%) collisions.  The ratio of the measured \ph meson \ptt distribution to the expected distribution is around unity and no significant difference is observed in central collisions, nor in peripheral collisions for $\ptt<2$~\gvc.  On the other hand, the average measured/expected ratio for the \ks is $0.6\pm0.1$ for $\ptt<3$~\gvc in central collisions, a deviation from unity of about four times larger than the uncertainties.  In peripheral collisions, the measured/expected ratio for the \ks does not appear to deviate significantly from unity for $\ptt<2$~\gvc.  For central \pb collisions, the shape of the \ptt distribution of \ks for $\ptt<3$~\gvc is consistent with the blast-wave parameterization of radial flow within uncertainties.  \linebreak Figure~\ref{fig:results:blast}(a) shows a \ks suppression of $\sim 40\%$ in the measured low-\ptt range and does not indicate the strongly momentum-dependent modification which is predicted by UrQMD for $\ptt<2$~\gvc~\cite{Bleicher_Stoecker,Bliecher_Aichelin}. However, this UrQMD calculation counts the suppression relative to the sum of both primary as well regenerated \ks resonances and therefore cannot be compared directly to the data.

The suggestion that \ks suppression does not have a strong \ptt dependence for $\ptt<3$~\gvc might be interpreted as evidence that the reduction observed in the \ks yield is not due to re-scattering.  However, it should be noted that regeneration is also expected to be more important at low \ptt, which could counteract some of the low-\ptt suppression that would be expected from re-scattering alone.  Furthermore, there is evidence for some increase in the measured/expected ratio for \ks from $\ptt=1.2$~\gvc to $\ptt=3$~\gvc for central collisions.  Further theoretical studies of the \ptt dependence of \ks suppression, with a full treatment of both re-scattering and regeneration, would be helpful in determining the likelihood that re-scattering is responsible for the observed decrease in the \ks yield.

\subsection{Mean Transverse Momentum\label{sec:results:mpt}}

\begin{figure*}
\includegraphics[width=38pc]{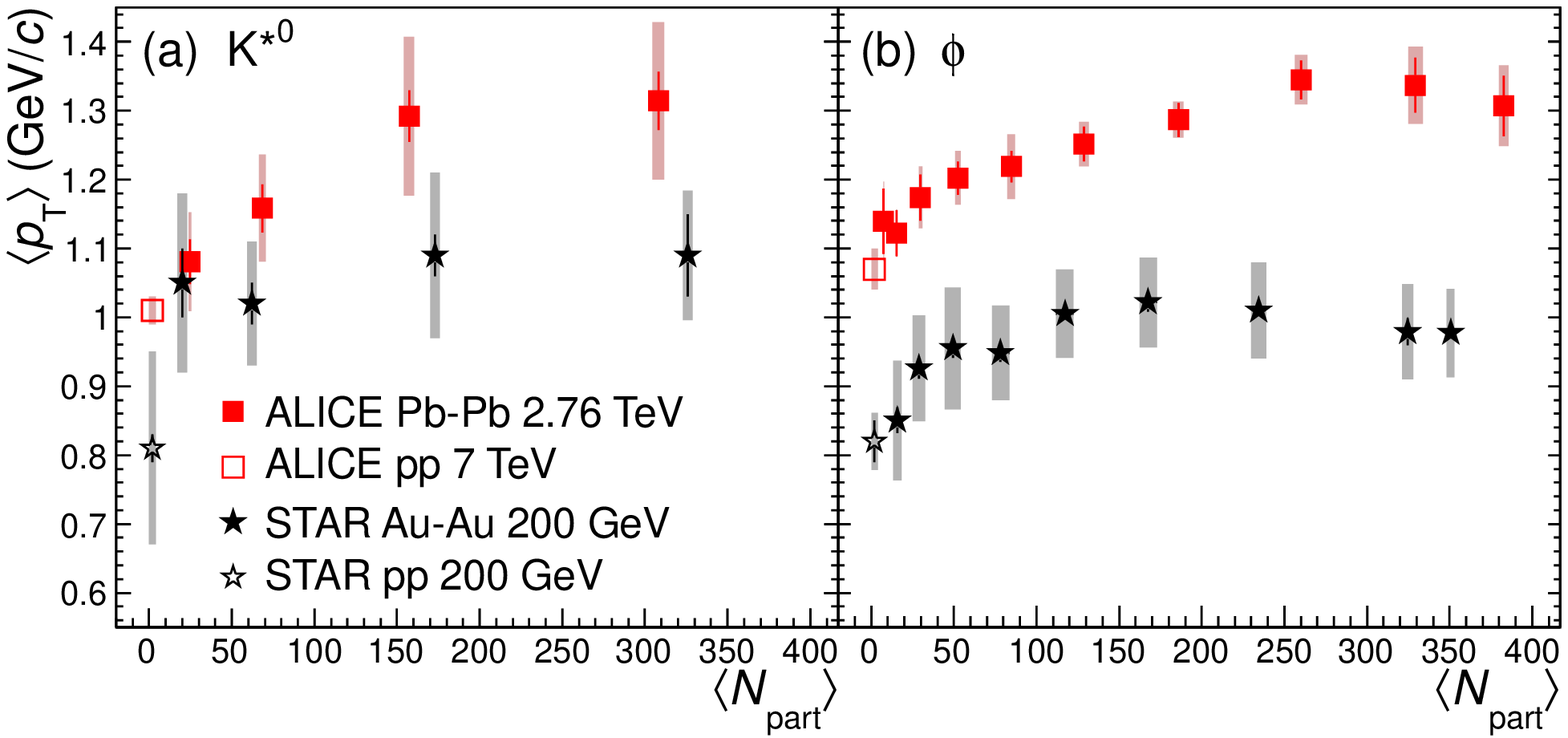}
\caption{\label{fig:results:mpt}(Color online)  Mean transverse momentum \mpt for \ks (a) and \ph mesons (b) as a function of \npart~\cite{ALICE_centrality} in \pb collisions at \rsnn.  Also shown are measurements of \mpt for minimum-bias pp collisions at \rs~\cite{ALICE_kstar_phi_7TeV} and \mpt for pp~\cite{STAR_Kstar_200GeV_2005,STAR_phi_200GeV_2005} and \au~\cite{STAR_Kstar_2011,STAR_phi_2009,STAR_centrality_2009} collisions at \rsnn[200~GeV].  Statistical uncertainties are shown as bars and systematic uncertainties are shown as boxes.  The most peripheral \ks points for \ada collisions (with $\npart\approx 20$) have been shifted horizontally for visibility.  The two most central \ph points for \au collisions are for overlapping centrality intervals (0-5\% and 0-10\%).}
\end{figure*}

The mean transverse momentum \mpt values for the \ks and \ph resonances are presented in Table~\ref{table:results} for different centrality intervals.  Figure~\ref{fig:results:mpt} shows \mpt for the \ks and \ph resonances in \pb collisions at \rsnn as a function of \npart~\cite{ALICE_centrality}.  Also shown are measurements of \mpt in pp collisions at \rs~\cite{ALICE_kstar_phi_7TeV} and in pp and \au collisions at \rsnn[200~GeV]~\cite{STAR_Kstar_200GeV_2005,STAR_Kstar_2011,STAR_phi_2009,STAR_phi_200GeV_2005,STAR_centrality_2009}.  The values of \mpt for the \ks (\ph) meson increase by about 20\% (15\%) in central \pb collisions relative to peripheral collisions.  The values of \mpt in pp collisions at \rs are consistent with the values observed in peripheral \pb collisions at \rsnn.  For central collisions, \mpt of the \ks (\ph) resonance measured in \pb collisions at \rsnn is about 20\% (30\%) higher than the values measured in \au collisions at \rsnn[200~GeV].  This is consistent with the observation~\cite{ALICE_piKp_PbPb} of increased radial flow in \ada collisions at the LHC relative to RHIC.

\begin{figure}
\includegraphics[width=19.1pc]{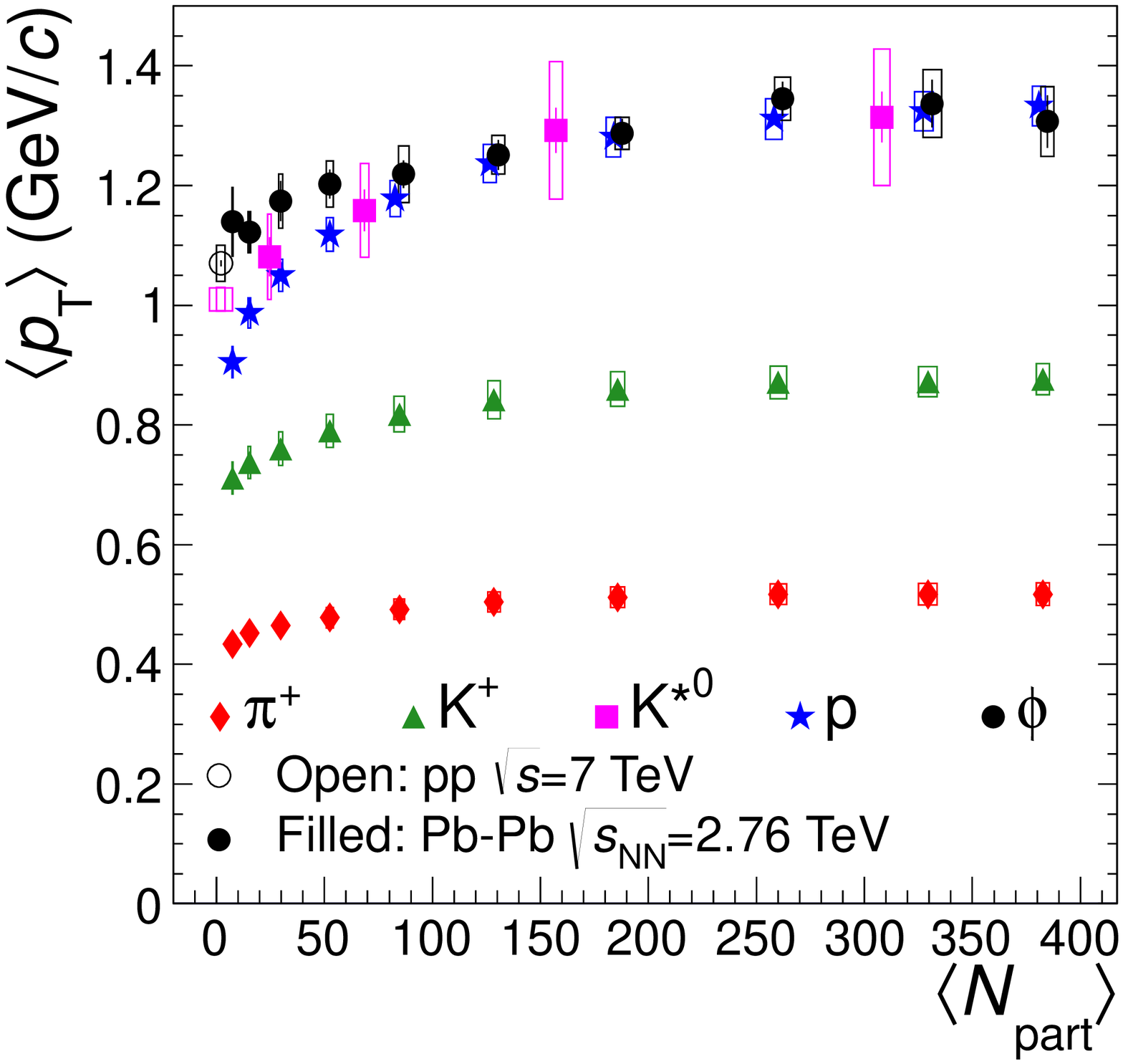}
\hspace{0.5pc}
\begin{minipage}{18pc}
\vspace{-20pc}
\caption{\label{fig:results:mpt_X}(Color online)  Mean transverse momentum of \pip, \kp, \ks, p, and \ph in \pb collisions at \rsnn (filled symbols)~\cite{ALICE_piKp_PbPb} as a function of \npart~\cite{ALICE_centrality}.  Also shown are \mpt values for the resonances in pp collisions at \rs~\cite{ALICE_kstar_phi_7TeV} (open symbols).  The measurements for central and mid-central \ph and p have been shifted horizontally for visibility.}
\end{minipage}
\end{figure}

The values of \mpt for \pip, \kp, \ks, p, and \ph in \pb collisions at \rsnn are shown in Fig.~\ref{fig:results:mpt_X} for different centrality intervals~\cite{ALICE_centrality,ALICE_piKp_PbPb}.  The values of \mpt for the resonances in pp collisions at \rs are also shown~\cite{ALICE_kstar_phi_7TeV}.  All particles exhibit an increase in \mpt from peripheral to central \pb collisions, but this increase is greatest for the protons.  While the increase in \mpt from the most peripheral to most central measured interval is about 20\% for \pip, \kp, \ks, and \ph, the value of \mpt for protons increases by 50\%.  For the 0-40\% most central collisions $(\npart\gtrsim100)$ the \mpt values of the \ks, proton, and \ph all appear to follow the same trend.  Within a given centrality interval, the \mpt values of these three particles are consistent with each other within uncertainties.  It should be noted that the masses of these three particles are similar: 896~\mvcc for the \ks, 938~\mvcc for the p, and 1019~\mvcc for the \ph~\cite{PDG}.  The similarity in \mpt values is consistent with expectations from a hydrodynamic framework, in which \ptt distributions would be determined predominantly by the particle masses.  This is discussed further in the context of the \ptt-dependent \pphi ratio in Sec.~\ref{sec:discussion:prod:phip}.

\subsection{Particle Production\label{sec:discussion:prod}}

In this section, the \ptt distributions and total yields of \ph mesons are compared to theoretical models and other particle species (with different baryon number, mass, or strange quark content) to study particle production mechanisms.  The \ph meson is used for these studies because it lives long enough that its yields and \ptt distributions do not appear to be affected by re-scattering or regeneration in the hadronic phase.  The possibility that such effects might change the \ks \ptt distributions and yields complicates any attempt to use that particle to study particle production.  The predictions of hydrodynamic models, which have described the yields and \ptt distributions of other hadrons with fair accuracy~\cite{ALICE_piKp_PbPb,ALICE_multistrange_PbPb,ALICE_multistrange_PbPb_erratum}, are compared in Sec.~\ref{sec:discussion:prod:theory} to the \ptt distribution of \ph mesons in central \pb collisions.  Differences in the production mechanisms of baryons and mesons can be studied through baryon-to-meson ratios.  The \ptt-dependent \omphi ratio, which compares baryons and mesons containing only strange (anti)quarks, is compared in Sec.~\ref{sec:discussion:prod:theory} to theoretical predictions and measurements in other collision systems.  If hadron production can be explained in a hydrodynamic framework, the particle mass plays an important role in determining the shape of the \ptt distribution.  To study this aspect of particle production, in Sec.~\ref{sec:discussion:prod:phip} \ptt distributions of \ph mesons are compared to protons, which are baryons with a different quark content but a very similar mass to the \ph.  The dependence of particle production on the strange quark content is explored in Sec.~\ref{sec:discussion:prod:enhancement}.  Here the enhancement of the \ph (which consists entirely of strange quarks but has no net strangeness) is compared to particles with 1, 2, and 3 units of open strangeness: the $\Lambda$, $\Xi$, and $\Omega$ baryons, respectively.

\subsubsection{Comparisons to Theoretical Models\label{sec:discussion:prod:theory}}

\begin{figure}[b!]
\includegraphics[width=19.1pc]{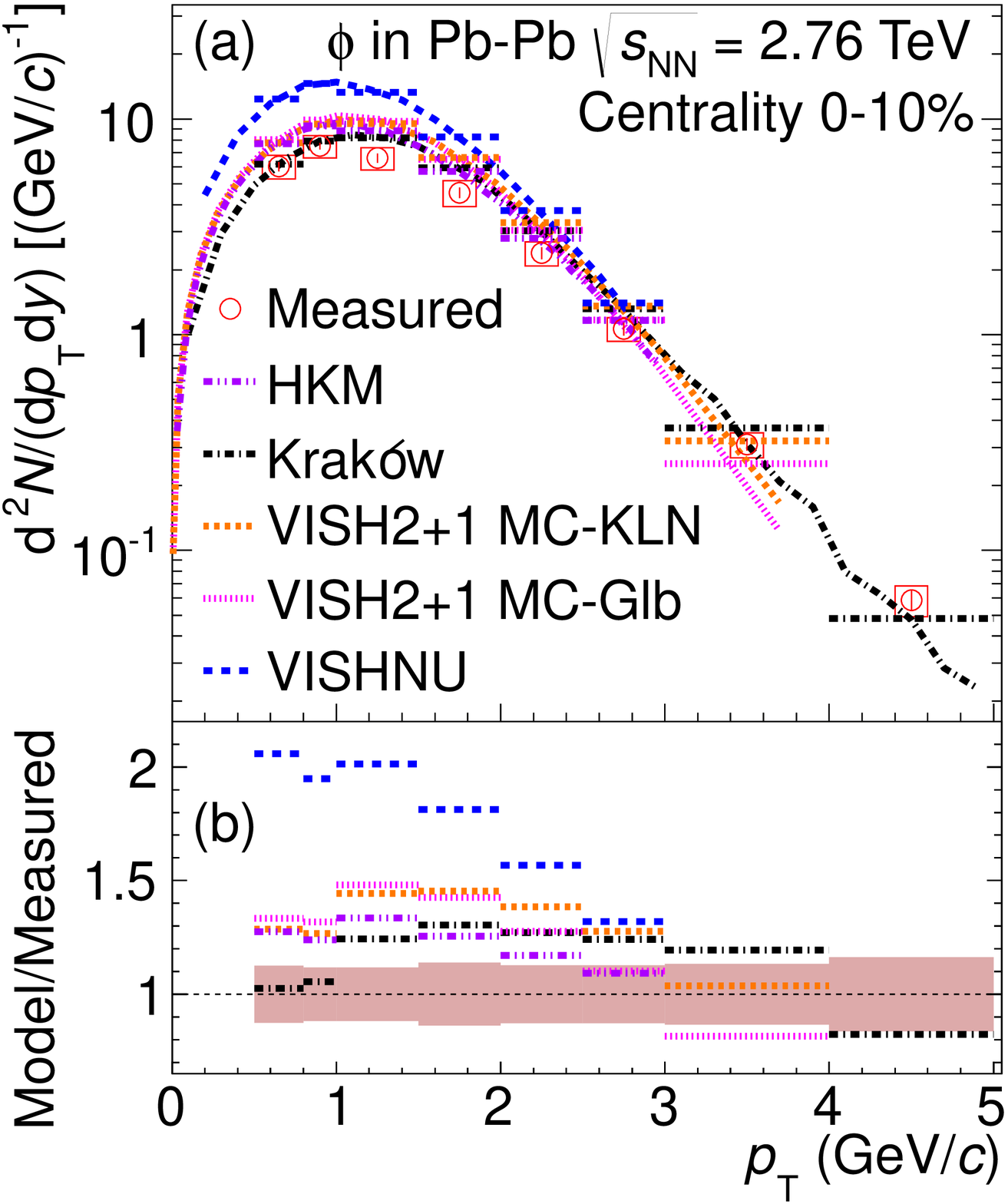}\hspace{0.5pc}
\begin{minipage}{18pc}
\vspace{-24pc}
\caption{\label{fig:results:theory}(Color online)  (a): Comparison of the measured \ph meson \ptt distribution in \pb collisions at \rsnn (centrality 0-10\%) to the distributions predicted by the Krak\'{o}w model~\cite{KRAKOW}, two versions of the VISH2+1 model~\cite{VISH2p1_MCGlb,VISH2p1_MCKLN}, the VISHNU~\cite{VISHNU_2013,VISHNU} model, and the HKM~\cite{HKM_2011,HKM_2013}.  The curves show the original predictions, while the horizontal lines show the predicted distributions re-binned so that they have compatible \ptt bins with the measured distribution. (b): The ratio of the re-binned predictions to the measured distribution for \ph mesons.  The shaded band shows the fractional uncertainty of the measured data points.}
\end{minipage}
\end{figure}

The measured \ph meson \ptt distribution for \pb collisions at \rsnn (centrality 0-10\%) is compared in Fig.~\ref{fig:results:theory} to five predicted distributions from hydrodynamic models.  The measured and predicted distributions are shown in panel (a), while the ratio of the predicted distributions to the \linebreak measured distribution is shown in panel (b).  VISH2+1 is a (2+1)-dimensional viscous hydrodynamic model~\cite{VISH2p1_MCGlb,VISH2p1_MCKLN}.  It has been observed to reproduce the total yields and the shapes of the \ptt distributions of \pion and K within about 25\% for $\ptt<2$~\gvc in central \pb collisions~\cite{ALICE_piKp_PbPb}.  VISH2+1 overestimates the total yields of the $\Xi$ and $\Omega$ baryons, though it provides a fair description of the shape of the $\Xi$ \ptt distribution~\cite{ALICE_multistrange_PbPb,ALICE_multistrange_PbPb_erratum}.  The VISH2+1 predictions shown in Fig.~\ref{fig:results:theory} are for two different sets of initial conditions: Monte-Carlo Kharzeev-Levin-Nardi initial conditions (MC-KLN) with $\eta/s=0.2$~\cite{VISH2p1_MCKLN} and Monte-Carlo Glauber (MC-Glb) initial conditions with $\eta/s=0.08$~\cite{VISH2p1_MCGlb}.  These predictions are larger than the measured \ph yield at low \ptt.  If the VISH2+1~MC-KLN prediction is fitted to the measured \ph data through multiplication by a \ptt-independent factor (0.74), it reproduces the shape of the measured \ptt distribution for $\ptt<3$ with a $\chi^{2}$ per degrees of freedom (\cn) value of 0.52 and no deviations beyond the experimental uncertainties.  Similarly, the VISH2+1~MC-Glb prediction can be fitted to the measured \ph \ptt distribution, with a multiplicative constant of 0.74, $\cn=1.1$, and no deviations beyond the experimental uncertainties.

The VISHNU model~\cite{VISHNU_2013,VISHNU} is a hybrid model which connects the VISH2+1 hydrodynamic description of the QGP to a microscopic hadronic transport model (UrQMD)~\cite{UrQMD_Bass,UrQMD} to describe the hadronic phase.  The VISNHU prediction for the \ph yield is larger than the measured data, and does not appear to reproduce the shape of the \ptt distribution (for $\ptt<3$~\gvc, $\cn=2.6$ when the prediction is fitted to the data with a multiplicative constant of 0.53).  It is also larger than either of the pure VISH2+1 predictions, which is attributed to the production of additional \ph mesons in the hadronic phase through $\km\kp$ scattering while \ph decays are turned off~\cite{VISHNU_2013}.

The hydrokinetic model (HKM)~\cite{HKM_2011,HKM_2013} combines an ideal hydrodynamic phase with a hadronic cascade (UrQMD) after the hydrodynamic description of the partonic phase.  Additional radial flow is built up during the hadronic phase and particle yields are affected by hadronic interactions (including baryon-antibaryon annihilation).  HKM has been observed to reproduce the measured \pion, K, proton, $\Xi$ and $\Omega$ data~\cite{ALICE_piKp_PbPb,ALICE_multistrange_PbPb,ALICE_multistrange_PbPb_erratum} better than VISH2+1, though it overestimates the yields of the multi-strange baryons.  The \ph yield is overestimated by HKM, though by a smaller amount than the VISH2+1 predictions.  The HKM prediction can be fitted to the measured \ph data through multiplication by a constant (0.80) for $\ptt<$~3~\gvc (its full range) with $\cn=0.53$ and no deviations beyond the experimental uncertainties.

The Krak\'{o}w model~\cite{KRAKOW} is a hydrodynamic model which introduces a bulk viscosity in the transition from the partonic to the hadronic phase, producing deviations from local equilibrium within the fluid elements, thereby affecting the hadron \ptt distributions and yields.  This model reproduces the \pion, K, and (anti)proton \ptt distributions within 20\% for $\ptt<3$~\gvc in central \pb collisions~\cite{ALICE_piKp_PbPb} and reproduces the $\Xi$ \ptt distributions within 30\% in the same \ptt range for the centrality range 0-60\%~\cite{ALICE_multistrange_PbPb,ALICE_multistrange_PbPb_erratum}.  It does not, however, describe the shape of the $\Omega$ \ptt distribution.  The Krak\'{o}w model over-predicts the \ph yield for $1<\ptt<4$~\gvc; however, it does not deviate from the measured yield by more than twice the uncertainty.  The Krak\'{o}w model prediction can be fitted to the measured \ph meson \ptt distribution through multiplication by a constant (0.85) for $\ptt<4$~\gvc with $\cn=1.1$ and no deviations beyond the uncertainties.

The hydrodynamic models considered above describe the measured \ph meson \ptt distribution with varying degrees of success.  All of these models over-predict the \ph yield, while all except the Krak\'{o}w model predict softer \ptt distributions for the \ph meson than was measured.  The best descriptions of the shape of the \ph meson \ptt distribution are given by the HKM and the Krak\'{o}w model.  Coupling hydrodynamics to a hadronic cascade, as is done in the KHM and VISHNU, has produced widely different results.  For the \ph, the two implementations of the VISH2+1 model produce similar results for $\ptt<2$~\gvc, despite having different initial conditions and viscosities.

\begin{figure}[h]
\includegraphics[width=19.1pc]{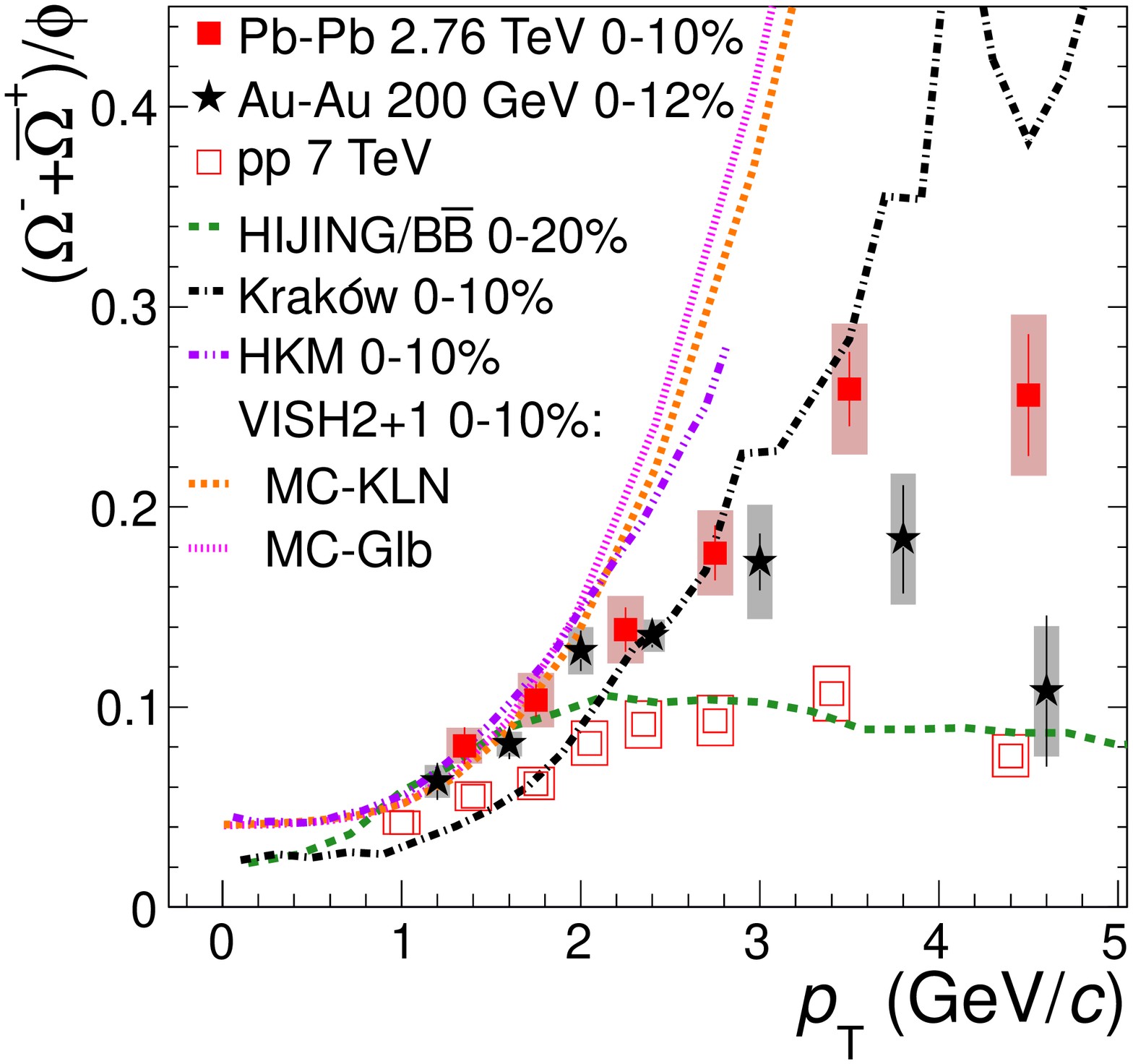}
\hspace{0.5pc}
\begin{minipage}{18pc}
\vspace{-19.2pc}
\caption{\label{fig:results:Omega2phi_pt}(Color online)  Ratio \omxphi as a function of \ptt for \pb collisions at \rsnn (centrality 0-10\%)~\cite{ALICE_multistrange_PbPb,ALICE_multistrange_PbPb_erratum}, pp collisions at \rs~\cite{ALICE_kstar_phi_7TeV,ALICE_multistrange_7TeV} and \au collisions at \rsnn[200~GeV] (centrality 0-12\%)~\cite{STAR_phi_2009}.  The statistical uncertainties are shown as bars, systematic uncertainties (including \ptt-uncorrelated and \ptt-correlated components) are shown as shaded boxes, and the sum in quadrature of the statistical and systematic uncertainties (for the pp data) is shown as open boxes.  Also shown are predictions of this ratio made by various models for central \pb collisions at \rsnn (centrality 0-20\% for HIJING/B$\overline{\mathrm{B}}$, centrality 0-10\% for the other models)~\cite{VISH2p1_MCGlb,VISH2p1_MCKLN,KRAKOW,HKM_2011,HKM_2013,HIJINGBBbar_2011a,HIJINGBBbar_2011b}.}
\end{minipage}
\end{figure}

The \ph and $\Omega$ are, respectively, a meson and a baryon made up entirely of strange (anti)quarks.  In some particle production models, such as the HIJING/B$\overline{\mathrm{B}}$ model~\cite{HIJINGBBbar_2011a,HIJINGBBbar_2011b}, soft particles are produced through string fragmentation.  The string tension is predicted~\cite{HIJINGBBbar_2011a} to influence the yields of strange particles, with multi-strange baryons and the \omphi ratio being particularly sensitive to the tension~\cite{LHC_predictions_2008}.  Figure~\ref{fig:results:Omega2phi_pt} shows the $\omphi\equiv\omxphi$ ratio as a function of \ptt in \pb collisions at \rsnn (centrality 0-10\%)~\cite{ALICE_multistrange_PbPb,ALICE_multistrange_PbPb_erratum}, pp collisions at \rs~\cite{ALICE_kstar_phi_7TeV,ALICE_multistrange_7TeV}, and \au collisions at \rsnn[200~GeV] (centrality 0-12\%)~\cite{STAR_phi_2009}.  The ratio measured in \pb collisions at \rsnn is consistent with the ratio measured in \au collisions at \rsnn[200~GeV] for $\ptt\lesssim 3$~\gvc, but is larger than the \au measurement at high \ptt.  Predictions from the HIJING/B$\overline{\mathrm{B}}$ and hydrodynamic models are also shown.  None of these models is able to predict the measured \omphi ratio.  HKM provides a better description of the $\Omega$ \ptt distributions than VISH2+1; however, it overestimates the total yield~\cite{ALICE_multistrange_PbPb,ALICE_multistrange_PbPb_erratum}.  The VISH2+1 and HKM predictions are consistent with the measured \omphi ratio for $\ptt<2$~\gvc, but increase faster with \ptt than the data for $\ptt>2$~\gvc.  The HKM does appear to provide a  better description of the slope of the measured \omphi ratio.  The Krak\'{o}w model under-predicts the measured data at low \ptt, but is consistent with the data for $2<\ptt<3.5$~\gvc.  This model is able to reproduce the measured $\Omega$ yield within about 30\%~\cite{ALICE_multistrange_PbPb,ALICE_multistrange_PbPb_erratum}, but does not reproduce the shape of the \ptt distribution.  The \omphi ratio predicted by the HIJING/B$\overline{\mathrm{B}}$ v2.0 model~\cite{HIJINGBBbar_2011a,HIJINGBBbar_2011b,HIJINGBBbar_2012,HIJINGBBbar_2014}, with a strong color field and a string tension of $\kappa=1.8$~GeV/fm, reproduces neither the shape nor the values of the measured data.  A larger string tension of $\kappa=5.1$~GeV/fm gives a predicted \omphi ratio (not shown) that is at least a factor of three larger than the measured ratio.  The same model can reproduce the \omphi ratio observed in pp collisions at \rs~\cite{ALICE_kstar_phi_7TeV,ALICE_multistrange_7TeV}\footnote{The prediction was calculated for \rs[5~TeV].} with a string tension of $\kappa=2$~GeV/fm, and describes the \omphi ratio observed in \au collisions at \rsnn[200~GeV]~\cite{LHC_predictions_2008,STAR_phi_200GeV_2006} with a string tension of $\kappa=3$~GeV/fm.

\begin{figure*}[h]
\includegraphics[width=38pc]{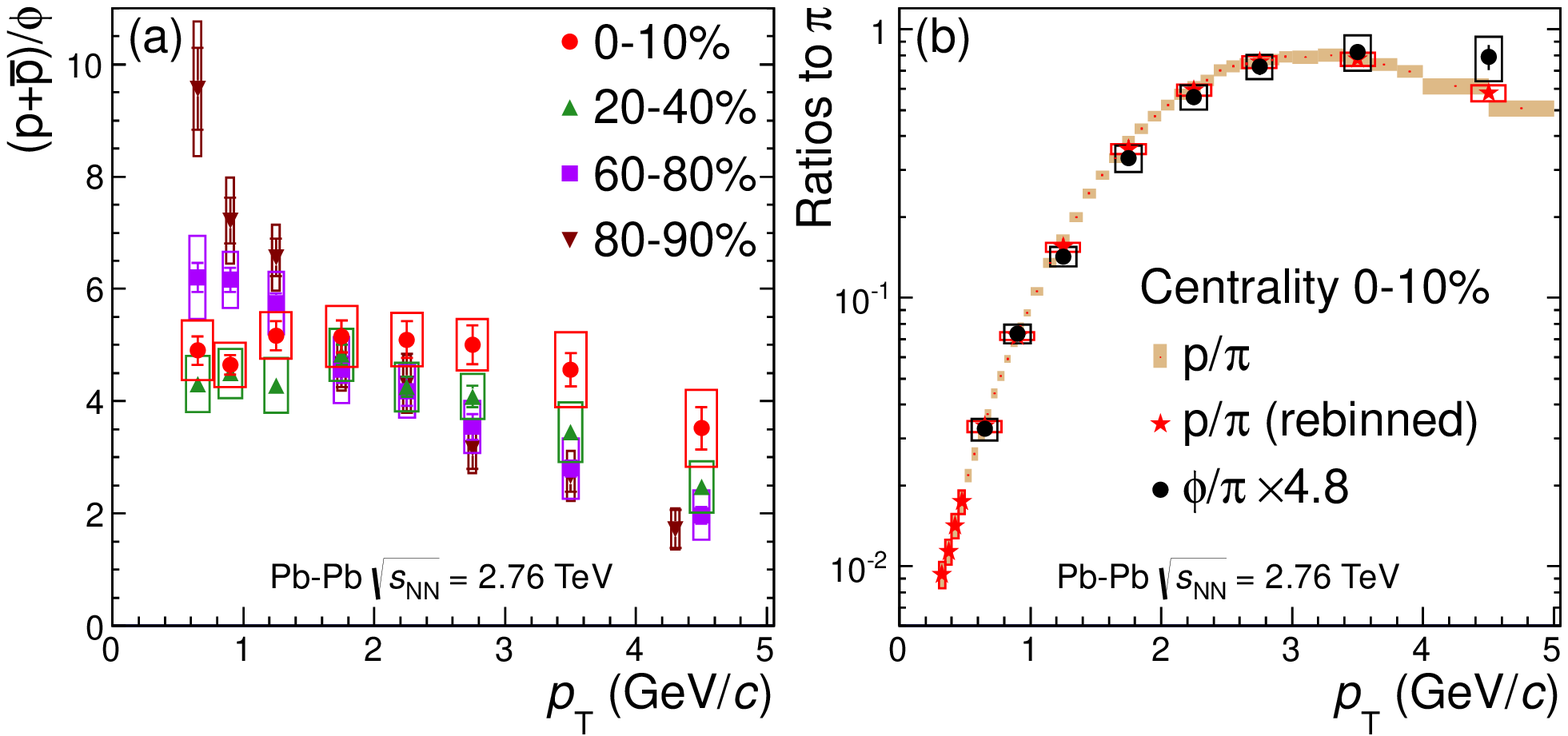}
\caption{\label{fig:results:ratios_pt}(Color online)  (a): Ratio \pphi as a function of \ptt for \pb collisions at \rsnn~\cite{ALICE_piKp_PbPb,ALICE_piKp_PbPb_combined} for four centrality intervals. (b): Ratios of p and \ph yields to charged pions as a function of \ptt for central \pb collisions at \rsnn~\cite{ALICE_piKp_PbPb,ALICE_piKp_PbPb_combined}.  The \ppi ratio is presented using two \ptt binning schemes: the ratio with its original measured bins is shown along with a recalculated version that uses the same bins as the \ph meson \ptt distribution for $0.5<\ptt<5$~\gvc.  In order to show the similarity of the shapes of the two ratios for $\ptt<3$~\gvc, the \phipi ratio has been scaled so that the \ph and proton integrated yields are identical.  In both panels, the statistical uncertainties are shown as bars and the total systematic uncertainties (including \ptt-uncorrelated and \ptt-correlated components) are shown as boxes.}
\end{figure*}

\subsubsection{Particles with Similar Masses\label{sec:discussion:prod:phip}}

The proton and \ph have similar masses, but different baryon numbers and quark content.  If production of these particles is described within a hydrodynamic framework, the \ptt distributions of these species are expected to have similar shapes, despite their different quantum numbers.  Figure~\ref{fig:results:ratios_pt}(a) shows the ratio $\pphi\equiv\pxphi$ as a function of \ptt for \pb collisions at \rsnn~\cite{ALICE_piKp_PbPb,ALICE_piKp_PbPb_combined} for different centralities.\footnote{The values of the \pphi ratio for the 10-20\% and 40-60\% centrality intervals, which are not shown here, are available in the Durham Reaction Database.}  For central collisions, the \pphi ratio is flat over the entire measured range.  However, for non-central collisions, this ratio is observed to decrease with \ptt.  This behavior can also be seen in the \mpt values of p and \ph in Fig.~\ref{fig:results:mpt_X}: these values are consistent with each other in central collisions, but \mpt is lower for p than for \ph in peripheral collisions.  The flat \pphi ratio in central collisions indicates that, at LHC energies, the shapes of the \ptt distributions of the p and \ph at low and intermediate \ptt are determined by the particle masses.  One possible explanation for the non-constant \pphi ratio in peripheral \pb collisions would be that the particles have a production mechanism in which the quark content is an important factor in determining the shapes of the \ptt distributions.  At RHIC energies, a splitting in the nuclear modification factor \rcp (the ratio of central to peripheral particle yields scaled by the number of binary collisions in the two centrality intervals), with baryons being less suppressed than mesons at intermediate \ptt~\cite{STAR_Kstar_200GeV_2005,STAR_pip_2006,STAR_K0sLambda_2004,STAR_phi_2007,STAR_XiOmega_2007}, has been taken as evidence in favor of recombination models~\cite{Coalescence_Review_2008}.  However, at LHC energies, the flat \pphi ratio suggests that recombination might not be suited to explain the shapes of the observed particle \ptt distributions in central \ada collisions at low and intermediate \ptt.

The $\ppi\equiv\pxpix$ ratio~\cite{ALICE_piKp_PbPb,ALICE_piKp_PbPb_combined} is shown in Fig.~\ref{fig:results:ratios_pt}(b).  When this ratio was first reported~\cite{ALICE_piKp_PbPb}, it was not clear if the observed increase in \ppi with transverse momentum is due to hydrodynamic effects or quark recombination.  As shown in Fig.~\ref{fig:results:ratios_pt}, the baryon-to-meson ratio \ppi has a very similar shape to the meson-to-meson ratio $\phipi\equiv\phipix$ for $\ptt<3$~\gvc.  This indicates that the number of quarks is not the main factor that determines the shapes of particle \ptt distributions at low and intermediate \ptt in central collisions.  This is contrary to the expectations from recombination, but consistent with hydrodynamic models.

\subsubsection{Strangeness Content\label{sec:discussion:prod:enhancement}}

The enhancement ratio is defined as the yield (\dndy) of a particle in \ada collisions normalized to \npart and divided by the same quantity in pp collisions\footnote{Reference yields measured in p--Be collisions have also been used~\cite{NA57_SE_40AGeV,NA57_SE_158AGeV}.} at the same energy.  This ratio has been the traditional way of presenting strangeness production in heavy-ion collisions~\cite{ALICE_multistrange_PbPb,ALICE_multistrange_PbPb_erratum,STAR_XiOmega_2007,NA57_SE_40AGeV,NA57_SE_158AGeV,WA97_enhancement,NA49_Omega_2005,NA49_Lambda_Xi_2008,NA49_Lambda_Xi_2009,STAR_multistrange_2004,STAR_SE_200GeV}.  However, given the fact that charged-particle production increases in a non-linear way with the number of participants~\cite{ALICE_multiplicity}, part of the enhancement observed using this ratio cannot be attributed to strangeness.  A way to avoid this bias is to normalize to the pion yield.  In order to allow for an easy comparison to previous measurements both approaches are discussed in this section.

The \ph yield in pp collisions at \rs[2.76~TeV] has been estimated by interpolating between the measured yields at \rs[900~GeV]~\cite{ALICE_strange_900GeV} and \rs~\cite{ALICE_kstar_phi_7TeV}, assuming that the yield varies as $s^{n}$.  Given the measured \ph yields at \rs[900~GeV] and \rs, the value of the power $n$ was found to be 0.10.  For comparison, the calculation of the enhancement values of multi-strange baryons at \rsnn uses an energy dependence of $s^{0.13}$ to find the interpolated pp reference values~\cite{ALICE_multistrange_PbPb,ALICE_multistrange_PbPb_erratum}.  The charged-particle pseudorapidity density is observed to vary as $s^{0.11}$~\cite{ALICE_multiplicity_PbPb}.  The systematic uncertainty in the interpolated \ph yield is estimated by successively increasing, then decreasing each of the two measured points by its own uncertainty and repeating the interpolation procedure.  The resulting variations in the interpolated yield are incorporated into the systematic uncertainty.  The systematic uncertainty in the interpolated \ph reference yield is 13\%.  Including the \ph meson yields measured in pp collisions at \rs[200~GeV]~\cite{STAR_phi_200GeV_2005,PHENIX_mesons_pp_2011} in the interpolation does not significantly alter the result.  The $\Lambda$ enhancement in \pb collisions at \rsnn, calculated using the yields reported in Ref.~\cite{ALICE_k0s_Lambda_PbPb}, is also reported below for the purpose of comparison with the \ph.  The reference $\Lambda$ yield in pp collisions at \rs[2.76~TeV] is estimated by extrapolating from the measured yield in (inelastic) pp collisions at \rs[900~GeV]~\cite{ALICE_strange_900GeV}, assuming the same energy dependence as \dnc.  The systematic uncertainty in this extrapolation is estimated by using the energy dependence of the $\Lambda+\overline{\Lambda}$ yield in non-single-diffractive pp collisions at \rs[200~GeV], 900~GeV, and 7~TeV~\cite{STAR_strange_pp_2007,CMS_strange_900GeV_7TeV}.  The uncertainty in the extrapolated $\Lambda$ reference yield is 19\%.

\begin{figure*}[t]
\includegraphics[width=38pc]{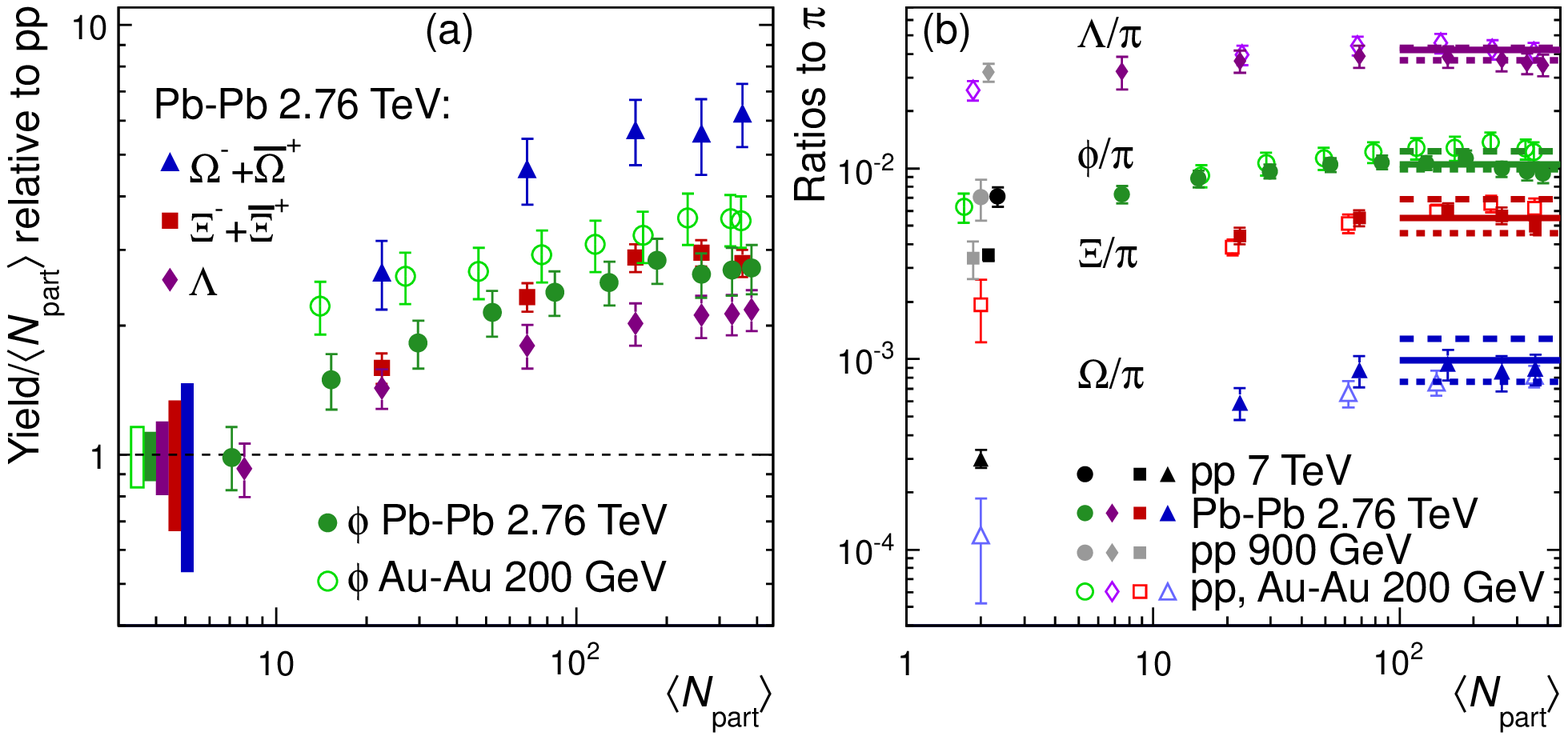}
\caption{\label{fig:results:se}(Color online)  (a): Enhancement of \ph, $\Lambda$, $\Xi$, and $\Omega$ in \pb collisions at \rsnn~\cite{ALICE_multistrange_PbPb,ALICE_multistrange_PbPb_erratum,ALICE_k0s_Lambda_PbPb}, calculated using pp reference yields (extrapolated for $\Lambda$, interpolated for \ph, $\Xi$, and $\Omega$).  Also shown is the enhancement of the \ph in \au collisions at \rsnn[200~GeV]~\cite{STAR_phi_2009b}, calculated using a measured pp reference yield.  The enhancement value reported here is the ratio of the yield (\dndy) of a particle in \ada collisions to the yield of that particle in pp collisions at the same energy, with both the numerator and denominator normalized by \npart.  The two most central \ph enhancement values for \au collisions are for overlapping centrality intervals (0-5\% and 0-10\%).  Bars represent the uncertainties in the \ada yields (including centrality-uncorrelated and centrality-correlated components), while the boxes at low values of \npart represent the uncertainties of the pp reference yields.  The \ph and $\Lambda$ measurements at $\npart=7.5$ have been shifted horizontally for visibility. (b):~Ratios of particle yields to charged pion yields for \pb collisions at \rsnn~\cite{ALICE_piKp_PbPb,ALICE_multistrange_PbPb,ALICE_multistrange_PbPb_erratum,ALICE_k0s_Lambda_PbPb}, \au collisions at \rsnn[200~GeV]~\cite{STAR_phi_2009,STAR_centrality_2009,STAR_XiOmega_2007,STAR_strange_2012}, and pp collisions at \rs[200~GeV]~\cite{STAR_strange_pp_2007,STAR_centrality_2009,STAR_phi_200GeV_2005}, 900~GeV~\cite{ALICE_strange_900GeV,ALICE_piKp_900GeV}, and 7 TeV~\cite{ALICE_kstar_phi_7TeV,ALICE_piKp_7TeV,ALICE_multistrange_7TeV}.  The lines show ratios given by grand-canonical thermal models with temperatures of 170~MeV~\cite{Cleymans_2006} (upper dashed lines), 164~MeV~\cite{Andronic2009,Andronic2009_Erratum} (solid lines), and 156~MeV~\cite{Stachel_SQM2013} (lower dashed lines).  The total uncertainties (including centrality-uncorrelated and centrality-correlated components) are shown as bars.  Some of the measurements at $\npart=2$ have been shifted horizontally for visibility.  The two most central \phipi values for \au collisions are for overlapping centrality intervals (0-5\% and 0-10\%).}
\end{figure*}

The enhancement values for \ph for different centrality intervals are shown in Fig.~\ref{fig:results:se}(a) along with the enhancement values for $\Lambda$, $\Xi$, and $\Omega$~\cite{ALICE_multistrange_PbPb,ALICE_multistrange_PbPb_erratum,ALICE_k0s_Lambda_PbPb}.  Enhancement values for \ph in \au collisions at \rsnn[200~GeV]~\cite{STAR_phi_2009b} are also shown.  The \ph enhancement ratio decreases from \rsnn[200~GeV] to \rsnn, a trend that has been observed for the other particles as well~\cite{ALICE_multistrange_PbPb,ALICE_multistrange_PbPb_erratum,STAR_SE_200GeV}.  The values of the \ph and $\Lambda$ enhancement for the 80-90\% centrality interval ($\npart=7.5$) are consistent with unity, \textit{i.e.}, the yields per participant nucleon of these particles in peripheral \pb collisions are consistent with the estimated yields in pp collisions.  The yields of \ph, $\Lambda$, $\Xi$, and $\Omega$ at LHC energies increase faster than linearly with \npart until $\npart\approx 100$, while the enhancement values seem to be saturated for higher values of \npart.  The enhancement values increase with the number of strange valence (anti)quarks, a trend which is also observed at lower energies.  For collisions at \rsnn, the \ph enhancement is consistent with the enhancement values of $\Lambda$ (one strange valence quark), as well as $\Xi^{-}$ and $\overline{\Xi}^{+}$ (two strange valence quarks or antiquarks).  The central values of the \ph enhancement tend to be between the $\Lambda$ and $\Xi$ enhancement values.  A similar behavior is observed when the \ph is compared to $\Lambda$, $\overline{\Lambda}$, and $\overline{\Xi}^{+}$ in \au collisions at \rsnn[200~GeV]~\cite{STAR_SE_200GeV,STAR_phi_2009b}.

As an alternative to the standard \npart-based enhancement ratio discussed above, the yields of particles containing strange quarks are compared to pion yields.  This is shown in Fig.~\ref{fig:results:se}(b) for \ada and pp collisions at RHIC and LHC energies~\cite{ALICE_piKp_PbPb,ALICE_strange_900GeV,STAR_strange_pp_2007,STAR_phi_2009,ALICE_kstar_phi_7TeV,ALICE_piKp_7TeV,STAR_centrality_2009,STAR_phi_200GeV_2005,ALICE_piKp_900GeV,ALICE_multistrange_PbPb,ALICE_multistrange_PbPb_erratum,ALICE_multistrange_7TeV,STAR_XiOmega_2007,ALICE_k0s_Lambda_PbPb}.  The ratios shown are $\phipi\equiv\ph/(\pim+\pip)$, $\xipi\equiv(\Xi^{-}+\overline{\Xi}^{+})/(\pim+\pip)$, and $\ompi\equiv(\Omega^{-}+\overline{\Omega}^{+})/(\pim+\pip)$.  For \pb collisions at \rsnn, $\lpi\equiv 2\Lambda/(\pim+\pip)$, but otherwise $\lpi\equiv(\Lambda+\overline{\Lambda})/(\pim+\pip)$.  While the \lpi,  \xipi, and \ompi ratios in pp collisions are higher at LHC energies than at RHIC energies, the \phipi ratio in pp collisions does not exhibit a significant change from \rs[200~GeV] to 7~TeV.  Relative to pp collisions, strangeness production in \pb collisions at \rsnn at first increases with centrality and appears to saturate for $\npart\gtrsim 100$.  A decrease in the \phipi, \lpi, and \xipi ratios for the 0-20\% most central collisions ($\npart\gtrsim 230$) may be present; however, the trend is flat within systematic uncertainties.  The increase in these ratios from pp to central \pb collisions at LHC energies is $\sim 3.3$ for \ompi, $\sim 1.6$ for \xipi, $\sim 1.2$ for \lpi, and $\sim 1.4$ for \phipi.  These values are about one half of the enhancement ratios discussed above.  The fractional increase in the \phipi ratio is similar to the increases observed in both the \lpi and \xipi ratios, a trend which is also observed in the standard enhancement ratios presented in the previous paragraph.  At SPS energies, a study of the \phipi, $\mathrm{K}\kern-0.05em/\krr\pion$, and $(\mathrm{K}\kern-0.05em/\krr\pion)^{2}$ ratios suggests that the \ph behaves as a particle with an effective strangeness quantum number between 1 and 2~\cite{Kraus_SPS_2007}.  The values of the \phipi, \lpi, \xipi, and \ompi ratios obtained from grand-canonical thermal models with temperatures of 170~MeV~\cite{Cleymans_2006} (upper dashed lines), 164~MeV~\cite{Andronic2009,Andronic2009_Erratum} (solid lines), and 156~MeV~\cite{Stachel_SQM2013} (lower dashed lines) are also shown.  It should be noted that the model using a temperature of 164~MeV gives a \ppi ratio that is about 50\% greater than the measured value.

\section{Conclusions\label{sec:conclusions}}

The \ptt distributions, masses, and widths of \ks and \ph mesons have been measured at mid-rapidity ($|y|<0.5$) in \pb collisions at \rsnn using the ALICE detector.  The masses and widths of these resonances, reconstructed via their hadronic decays, are consistent with the vacuum values.  The measured \mpt is 15-20\% higher in central \pb collisions than in peripheral collisions and it is found to be higher in \ada collisions at LHC energies than at RHIC energies.  This suggests stronger radial flow at the LHC, which has also been concluded based on previous measurements of pion, kaon, and proton \ptt distributions.  Relative to the yields of charged kaons, the total yield (\dndy) of \ks is observed to be suppressed in central \pb collisions.  When plotted as a function of \dncr, the \kskm ratio appears to follow a single trend for both RHIC and LHC energies and for different collision systems.  In contrast, no suppression is observed for the \ph.  When the \ptt distributions of the \ks and \ph mesons are compared to expected distributions based on the blast-wave model (using parameters taken from fits to other hadrons), \ks suppression is observed in central collisions for transverse momenta $\ptt<3$~\gvc.  The suppression of the integrated \ks yield might be taken to suggest that re-scattering of resonance decay products in the hadronic phase reduces the measurable yield of \ks mesons.  However, it is unclear if such a scenario can fully explain the observed \ptt dependence of the \ks suppression or the absence of broadening in its invariant-mass distribution.  The lack of suppression for the \ph meson could indicate that this particle decays outside the fireball due to its longer lifetime.  The measured \kskm ratio is compared to an extended thermal-model prediction~\cite{Markert_thermal,Torrieri_thermal,Torrieri_thermal_2001b,Torrieri_thermal_2001b_erratum} that includes re-scattering effects.  By assuming a chemical freeze-out temperature of 156~MeV, a model-dependent estimate of 2~fm/$c$ as the lower limit of the time between the chemical and kinetic freeze-out is extracted.  The measurement of at least one more resonance-to-stable ratio [such as $\Lambda(1520)\krl/\krr\Lambda$] will allow both the lifetime of the hadronic phase and the chemical freeze-out temperature to be estimated simultaneously within the framework of this model.  At LHC energies the \ph, which has hidden strangeness, is enhanced by an amount similar to particles with one or two units of open strangeness.  While a hydrodynamic framework can roughly describe the measured particle \ptt distributions in \pb collisions at \rsnn, inconsistencies nevertheless remain.  For central collisions the \pphi ratio is flat as a function of transverse momentum for $\ptt\lesssim 3$~\gvc.  This is consistent with hydrodynamic models, thereby suggesting that mass and hence radial flow plays a dominant role in the determination of the shapes of \ptt distributions at low and intermediate \ptt.  Models based on hydrodynamics (Krak\'{o}w, HKM, and VISH2+1) are able to reproduce the shape of the \ph meson \ptt distribution fairly well, but overestimate the \ph yield.  These models describe the \ptt distributions of other particles, such as \pion, K, and protons, reasonably well, but they encounter difficulties in describing the \ptt distribution of the $\Omega$ and the \omphi ratio.

\newenvironment{acknowledgement}{\relax}{\relax}
\begin{acknowledgement}
\section*{Acknowledgements}
The ALICE Collaboration would like to thank all its engineers and technicians for their invaluable contributions to the construction of the experiment and the CERN accelerator teams for the outstanding performance of the LHC complex.
%\\
The ALICE Collaboration gratefully acknowledges the resources and support provided by all Grid centres and the Worldwide LHC Computing Grid (WLCG) collaboration.
%\\
The ALICE Collaboration acknowledges the following funding agencies for their support in building and
running the ALICE detector:
 %\\
State Committee of Science,  World Federation of Scientists (WFS)
and Swiss Fonds Kidagan, Armenia,
 %\\
Conselho Nacional de Desenvolvimento Cient\'{\i}fico e Tecnol\'{o}gico (CNPq), Financiadora de Estudos e Projetos (FINEP),
Funda\c{c}\~{a}o de Amparo \`{a} Pesquisa do Estado de S\~{a}o Paulo (FAPESP);
 %\\
National Natural Science Foundation of China (NSFC), the Chinese Ministry of Education (CMOE)
and the Ministry of Science and Technology of China (MSTC);
 %\\
Ministry of Education and Youth of the Czech Republic;
 %\\
Danish Natural Science Research Council, the Carlsberg Foundation and the Danish National Research Foundation;
 %\\
The European Research Council under the European Community's Seventh Framework Programme;
 %\\
Helsinki Institute of Physics and the Academy of Finland;
 %\\
French CNRS-IN2P3, the `Region Pays de Loire', `Region Alsace', `Region Auvergne' and CEA, France;
 %\\
German BMBF and the Helmholtz Association;
%\\
General Secretariat for Research and Technology, Ministry of
Development, Greece;
%\\
Hungarian OTKA and National Office for Research and Technology (NKTH);
 %\\
Department of Atomic Energy and Department of Science and Technology of the Government of India;
 %\\
Istituto Nazionale di Fisica Nucleare (INFN) and Centro Fermi -
Museo Storico della Fisica e Centro Studi e Ricerche ``Enrico
Fermi", Italy;
 %\\
MEXT Grant-in-Aid for Specially Promoted Research, Ja\-pan;
 %\\
Joint Institute for Nuclear Research, Dubna;
 %\\
%Korea Foundation for International Cooperation of Science and Technology (KICOS);
National Research Foundation of Korea (NRF);
 %\\
CONACYT, DGAPA, M\'{e}xico, ALFA-EC and the EPLANET Program
(European Particle Physics Latin American Network)
 %\\
Stichting voor Fundamenteel Onderzoek der Materie (FOM) and the Nederlandse Organisatie voor Wetenschappelijk Onderzoek (NWO), Netherlands;
 %\\
Research Council of Norway (NFR);
 %\\
Polish Ministry of Science and Higher Education;
 %\\
National Science Centre, Poland;
 %\\
 Ministry of National Education/Institute for Atomic Physics and CNCS-UEFISCDI - Romania;
 %\\
Ministry of Education and Science of Russian Federation, Russian
Academy of Sciences, Russian Federal Agency of Atomic Energy,
Russian Federal Agency for Science and Innovations and The Russian
Foundation for Basic Research;
 %\\
Ministry of Education of Slovakia;
 %\\
Department of Science and Technology, South Africa;
 %\\
CIEMAT, EELA, Ministerio de Econom\'{i}a y Competitividad (MINECO) of Spain, Xunta de Galicia (Conseller\'{\i}a de Educaci\'{o}n),
CEA\-DEN, Cubaenerg\'{\i}a, Cuba, and IAEA (International Atomic Energy Agency);
 %\\
Swedish Research Council (VR) and Knut $\&$ Alice Wallenberg
Foundation (KAW);
 %\\
Ukraine Ministry of Education and Science;
 %\\
United Kingdom Science and Technology Facilities Council (STFC);
 %\\
The United States Department of Energy, the United States National
Science Foundation, the State of Texas, and the State of Ohio.
\end{acknowledgement}

\bibliographystyle{epj}
\bibliography{refs}

\newpage

\appendix
\section{The ALICE Collaboration}
\label{app:collab}

% Collaboration: CERN-LHC-ALICE
% Generation Date is 2014/Feb/20

% How to use:
%%%%%%%%% appendix with author list
%\appendix
%\section{The ALICE Collaboration}
%\label{app:collab}
%\input{authors-list.tex}  %%%%%%% get the latest version before submitting

\begingroup
\small
\begin{flushleft}
B.~Abelev\Irefn{org69}\And
J.~Adam\Irefn{org37}\And
D.~Adamov\'{a}\Irefn{org77}\And
M.M.~Aggarwal\Irefn{org81}\And
M.~Agnello\Irefn{org104}\textsuperscript{,}\Irefn{org87}\And
A.~Agostinelli\Irefn{org26}\And
N.~Agrawal\Irefn{org44}\And
Z.~Ahammed\Irefn{org123}\And
N.~Ahmad\Irefn{org18}\And
A.~Ahmad~Masoodi\Irefn{org18}
% ...
\renewcommand\labelenumi{\textsuperscript{\theenumi}~}

\section*{Affiliation notes}
\renewcommand\theenumi{\roman{enumi}}
\begin{Authlist}
\item \Adef{0}Deceased
\item \Adef{idp1075984}{Also at: St. Petersburg State Polytechnical University, St. Petersburg, Russia}
\item \Adef{idp2926016}{Also at: Department of Applied Physics, Aligarh Muslim University, Aligarh, India}
\item \Adef{idp3610448}{Also at: M.V. Lomonosov Moscow State University, D.V. Skobeltsyn Institute of Nuclear Physics, Moscow, Russia}
\item \Adef{idp3854048}{Also at: University of Belgrade, Faculty of Physics and ``Vin\v{c}a" Institute of Nuclear Sciences, Belgrade, Serbia}
\item \Adef{idp4134896}{Permanent Address: Permanent Address: Konkuk University, Seoul, Korea}
\item \Adef{idp4680704}{Also at: Institute of Theoretical Physics, University of Wroclaw, Wroclaw, Poland}
\item \Adef{idp5567136}{Also at: University of Kansas, Lawrence, KS, United States}
\end{Authlist}

\section*{Collaboration Institutes}
\renewcommand\theenumi{\arabic{enumi}~}
\begin{Authlist}

\item \Idef{org1}A.I. Alikhanyan National Science Laboratory (Yerevan Physics Institute) Foundation, Yerevan, Armenia
\item \Idef{org2}Benem\'{e}rita Universidad Aut\'{o}noma de Puebla, Puebla, Mexico
\item \Idef{org3}Bogolyubov Institute for Theoretical Physics, Kiev, Ukraine
\item \Idef{org4}Bose Institute, Department of Physics and Centre for Astroparticle Physics and Space Science (CAPSS), Kolkata, India
\item \Idef{org5}Budker Institute for Nuclear Physics, Novosibirsk, Russia
\item \Idef{org6}California Polytechnic State University, San Luis Obispo, CA, United States
\item \Idef{org7}Central China Normal University, Wuhan, China
\item \Idef{org8}Centre de Calcul de l'IN2P3, Villeurbanne, France
\item \Idef{org9}Centro de Aplicaciones Tecnol\'{o}gicas y Desarrollo Nuclear (CEADEN), Havana, Cuba
\item \Idef{org10}Centro de Investigaciones Energ\'{e}ticas Medioambientales y Tecnol\'{o}gicas (CIEMAT), Madrid, Spain
\item \Idef{org11}Centro de Investigaci\'{o}n y de Estudios Avanzados (CINVESTAV), Mexico City and M\'{e}rida, Mexico
\item \Idef{org12}Centro Fermi - Museo Storico della Fisica e Centro Studi e Ricerche ``Enrico Fermi'', Rome, Italy
\item \Idef{org13}Chicago State University, Chicago, USA
\item \Idef{org14}Commissariat \`{a} l'Energie Atomique, IRFU, Saclay, France
\item \Idef{org15}COMSATS Institute of Information Technology (CIIT), Islamabad, Pakistan
\item \Idef{org16}Departamento de F\'{\i}sica de Part\'{\i}culas and IGFAE, Universidad de Santiago de Compostela, Santiago de Compostela, Spain
\item \Idef{org17}Department of Physics and Technology, University of Bergen, Bergen, Norway
\item \Idef{org18}Department of Physics, Aligarh Muslim University, Aligarh, India
\item \Idef{org19}Department of Physics, Ohio State University, Columbus, OH, United States
\item \Idef{org20}Department of Physics, Sejong University, Seoul, South Korea
\item \Idef{org21}Department of Physics, University of Oslo, Oslo, Norway
\item \Idef{org22}Dipartimento di Fisica dell'Universit\`{a} `La Sapienza' and Sezione INFN Rome, Italy
\item \Idef{org23}Dipartimento di Fisica dell'Universit\`{a} and Sezione INFN, Cagliari, Italy
\item \Idef{org24}Dipartimento di Fisica dell'Universit\`{a} and Sezione INFN, Trieste, Italy
\item \Idef{org25}Dipartimento di Fisica dell'Universit\`{a} and Sezione INFN, Turin, Italy
\item \Idef{org26}Dipartimento di Fisica e Astronomia dell'Universit\`{a} and Sezione INFN, Bologna, Italy
\item \Idef{org27}Dipartimento di Fisica e Astronomia dell'Universit\`{a} and Sezione INFN, Catania, Italy
\item \Idef{org28}Dipartimento di Fisica e Astronomia dell'Universit\`{a} and Sezione INFN, Padova, Italy
\item \Idef{org29}Dipartimento di Fisica `E.R.~Caianiello' dell'Universit\`{a} and Gruppo Collegato INFN, Salerno, Italy
\item \Idef{org30}Dipartimento di Scienze e Innovazione Tecnologica dell'Universit\`{a} del  Piemonte Orientale and Gruppo Collegato INFN, Alessandria, Italy
\item \Idef{org31}Dipartimento Interateneo di Fisica `M.~Merlin' and Sezione INFN, Bari, Italy
\item \Idef{org32}Division of Experimental High Energy Physics, University of Lund, Lund, Sweden
\item \Idef{org33}Eberhard Karls Universit\"{a}t T\"{u}bingen, T\"{u}bingen, Germany
\item \Idef{org34}European Organization for Nuclear Research (CERN), Geneva, Switzerland
\item \Idef{org35}Faculty of Engineering, Bergen University College, Bergen, Norway
\item \Idef{org36}Faculty of Mathematics, Physics and Informatics, Comenius University, Bratislava, Slovakia
\item \Idef{org37}Faculty of Nuclear Sciences and Physical Engineering, Czech Technical University in Prague, Prague, Czech Republic
\item \Idef{org38}Faculty of Science, P.J.~\v{S}af\'{a}rik University, Ko\v{s}ice, Slovakia
\item \Idef{org39}Frankfurt Institute for Advanced Studies, Johann Wolfgang Goethe-Universit\"{a}t Frankfurt, Frankfurt, Germany
\item \Idef{org40}Gangneung-Wonju National University, Gangneung, South Korea
\item \Idef{org41}Gauhati University, Department of Physics, Guwahati, India
\item \Idef{org42}Helsinki Institute of Physics (HIP), Helsinki, Finland
\item \Idef{org43}Hiroshima University, Hiroshima, Japan
\item \Idef{org44}Indian Institute of Technology Bombay (IIT), Mumbai, India
\item \Idef{org45}Indian Institute of Technology Indore, Indore (IITI), India
\item \Idef{org46}Institut de Physique Nucl\'eaire d'Orsay (IPNO), Universit\'e Paris-Sud, CNRS-IN2P3, Orsay, France
\item \Idef{org47}Institut f\"{u}r Informatik, Johann Wolfgang Goethe-Universit\"{a}t Frankfurt, Frankfurt, Germany
\item \Idef{org48}Institut f\"{u}r Kernphysik, Johann Wolfgang Goethe-Universit\"{a}t Frankfurt, Frankfurt, Germany
\item \Idef{org49}Institut f\"{u}r Kernphysik, Westf\"{a}lische Wilhelms-Universit\"{a}t M\"{u}nster, M\"{u}nster, Germany
\item \Idef{org50}Institut Pluridisciplinaire Hubert Curien (IPHC), Universit\'{e} de Strasbourg, CNRS-IN2P3, Strasbourg, France
\item \Idef{org51}Institute for Nuclear Research, Academy of Sciences, Moscow, Russia
\item \Idef{org52}Institute for Subatomic Physics of Utrecht University, Utrecht, Netherlands
\item \Idef{org53}Institute for Theoretical and Experimental Physics, Moscow, Russia
\item \Idef{org54}Institute of Experimental Physics, Slovak Academy of Sciences, Ko\v{s}ice, Slovakia
\item \Idef{org55}Institute of Physics, Academy of Sciences of the Czech Republic, Prague, Czech Republic
\item \Idef{org56}Institute of Physics, Bhubaneswar, India
\item \Idef{org57}Institute of Space Science (ISS), Bucharest, Romania
\item \Idef{org58}Instituto de Ciencias Nucleares, Universidad Nacional Aut\'{o}noma de M\'{e}xico, Mexico City, Mexico
\item \Idef{org59}Instituto de F\'{\i}sica, Universidad Nacional Aut\'{o}noma de M\'{e}xico, Mexico City, Mexico
\item \Idef{org60}iThemba LABS, National Research Foundation, Somerset West, South Africa
\item \Idef{org61}Joint Institute for Nuclear Research (JINR), Dubna, Russia
\item \Idef{org62}Korea Institute of Science and Technology Information, Daejeon, South Korea
\item \Idef{org63}KTO Karatay University, Konya, Turkey
\item \Idef{org64}Laboratoire de Physique Corpusculaire (LPC), Clermont Universit\'{e}, Universit\'{e} Blaise Pascal, CNRS--IN2P3, Clermont-Ferrand, France
\item \Idef{org65}Laboratoire de Physique Subatomique et de Cosmologie, Universit\'{e} Grenoble-Alpes, CNRS-IN2P3, Grenoble, France
\item \Idef{org66}Laboratori Nazionali di Frascati, INFN, Frascati, Italy
\item \Idef{org67}Laboratori Nazionali di Legnaro, INFN, Legnaro, Italy
\item \Idef{org68}Lawrence Berkeley National Laboratory, Berkeley, CA, United States
\item \Idef{org69}Lawrence Livermore National Laboratory, Livermore, CA, United States
\item \Idef{org70}Moscow Engineering Physics Institute, Moscow, Russia
\item \Idef{org71}National Centre for Nuclear Studies, Warsaw, Poland
\item \Idef{org72}National Institute for Physics and Nuclear Engineering, Bucharest, Romania
\item \Idef{org73}National Institute of Science Education and Research, Bhubaneswar, India
\item \Idef{org74}Niels Bohr Institute, University of Copenhagen, Copenhagen, Denmark
\item \Idef{org75}Nikhef, National Institute for Subatomic Physics, Amsterdam, Netherlands
\item \Idef{org76}Nuclear Physics Group, STFC Daresbury Laboratory, Daresbury, United Kingdom
\item \Idef{org77}Nuclear Physics Institute, Academy of Sciences of the Czech Republic, \v{R}e\v{z} u Prahy, Czech Republic
\item \Idef{org78}Oak Ridge National Laboratory, Oak Ridge, TN, United States
\item \Idef{org79}Petersburg Nuclear Physics Institute, Gatchina, Russia
\item \Idef{org80}Physics Department, Creighton University, Omaha, NE, United States
\item \Idef{org81}Physics Department, Panjab University, Chandigarh, India
\item \Idef{org82}Physics Department, University of Athens, Athens, Greece
\item \Idef{org83}Physics Department, University of Cape Town, Cape Town, South Africa
\item \Idef{org84}Physics Department, University of Jammu, Jammu, India
\item \Idef{org85}Physics Department, University of Rajasthan, Jaipur, India
\item \Idef{org86}Physikalisches Institut, Ruprecht-Karls-Universit\"{a}t Heidelberg, Heidelberg, Germany
\item \Idef{org87}Politecnico di Torino, Turin, Italy
\item \Idef{org88}Purdue University, West Lafayette, IN, United States
\item \Idef{org89}Pusan National University, Pusan, South Korea
\item \Idef{org90}Research Division and ExtreMe Matter Institute EMMI, GSI Helmholtzzentrum f\"ur Schwerionenforschung, Darmstadt, Germany
\item \Idef{org91}Rudjer Bo\v{s}kovi\'{c} Institute, Zagreb, Croatia
\item \Idef{org92}Russian Federal Nuclear Center (VNIIEF), Sarov, Russia
\item \Idef{org93}Russian Research Centre Kurchatov Institute, Moscow, Russia
\item \Idef{org94}Saha Institute of Nuclear Physics, Kolkata, India
\item \Idef{org95}School of Physics and Astronomy, University of Birmingham, Birmingham, United Kingdom
\item \Idef{org96}Secci\'{o}n F\'{\i}sica, Departamento de Ciencias, Pontificia Universidad Cat\'{o}lica del Per\'{u}, Lima, Peru
\item \Idef{org97}Sezione INFN, Bari, Italy
\item \Idef{org98}Sezione INFN, Bologna, Italy
\item \Idef{org99}Sezione INFN, Cagliari, Italy
\item \Idef{org100}Sezione INFN, Catania, Italy
\item \Idef{org101}Sezione INFN, Padova, Italy
\item \Idef{org102}Sezione INFN, Rome, Italy
\item \Idef{org103}Sezione INFN, Trieste, Italy
\item \Idef{org104}Sezione INFN, Turin, Italy
\item \Idef{org105}SSC IHEP of NRC Kurchatov institute, Protvino, Russia
\item \Idef{org106}SUBATECH, Ecole des Mines de Nantes, Universit\'{e} de Nantes, CNRS-IN2P3, Nantes, France
\item \Idef{org107}Suranaree University of Technology, Nakhon Ratchasima, Thailand
\item \Idef{org108}Technical University of Split FESB, Split, Croatia
\item \Idef{org109}The Henryk Niewodniczanski Institute of Nuclear Physics, Polish Academy of Sciences, Cracow, Poland
\item \Idef{org110}The University of Texas at Austin, Physics Department, Austin, TX, USA
\item \Idef{org111}Universidad Aut\'{o}noma de Sinaloa, Culiac\'{a}n, Mexico
\item \Idef{org112}Universidade de S\~{a}o Paulo (USP), S\~{a}o Paulo, Brazil
\item \Idef{org113}Universidade Estadual de Campinas (UNICAMP), Campinas, Brazil
\item \Idef{org114}University of Houston, Houston, TX, United States
\item \Idef{org115}University of Jyv\"{a}skyl\"{a}, Jyv\"{a}skyl\"{a}, Finland
\item \Idef{org116}University of Liverpool, Liverpool, United Kingdom
\item \Idef{org117}University of Tennessee, Knoxville, TN, United States
\item \Idef{org118}University of Tokyo, Tokyo, Japan
\item \Idef{org119}University of Tsukuba, Tsukuba, Japan
\item \Idef{org120}University of Zagreb, Zagreb, Croatia
\item \Idef{org121}Universit\'{e} de Lyon, Universit\'{e} Lyon 1, CNRS/IN2P3, IPN-Lyon, Villeurbanne, France
\item \Idef{org122}V.~Fock Institute for Physics, St. Petersburg State University, St. Petersburg, Russia
\item \Idef{org123}Variable Energy Cyclotron Centre, Kolkata, India
\item \Idef{org124}Vestfold University College, Tonsberg, Norway
\item \Idef{org125}Warsaw University of Technology, Warsaw, Poland
\item \Idef{org126}Wayne State University, Detroit, MI, United States
\item \Idef{org127}Wigner Research Centre for Physics, Hungarian Academy of Sciences, Budapest, Hungary
\item \Idef{org128}Yale University, New Haven, CT, United States
\item \Idef{org129}Yonsei University, Seoul, South Korea
\item \Idef{org130}Zentrum f\"{u}r Technologietransfer und Telekommunikation (ZTT), Fachhochschule Worms, Worms, Germany
\end{Authlist}
\endgroup

\end{document}